\newcommand*\dash{\unskip\kern.16667em---\penalty\exhyphenpenalty
        \hskip.16667em\relax
}

\documentclass[letterpaper,twocolumn,10pt]{article}
\usepackage{usenix-2020-09}
\usepackage{microtype} 
\usepackage{minipage}
\usepackage{booktabs}
\usepackage{xltabular}
\usepackage{tabularx}
\usepackage{xspace}
\usepackage[numbers]{natbib}
\usepackage{adjustbox} 

\usepackage{colortbl}

\usepackage{amsmath}

\usepackage{enumitem}

\usepackage{hyperref}

\usepackage{tcolorbox}
 
\usepackage{comment}
\definecolor{Red}{rgb}{1,0,0}
\definecolor{Green}{rgb}{0,0.8,0.5}
\definecolor{Purple}{rgb}{0.75,0,1}
\definecolor{Orange}{rgb}{255, 87, 51}
\definecolor{babypink}{rgb}{0.96, 0.76, 0.76}
\definecolor{azure}{rgb}{0,0.49,1}
\definecolor{periwinkle}{rgb}{0.8, 0.8, 1.0}
\definecolor{Pink}{RGB}{255, 102, 204}
\definecolor{electriccyan}{rgb}{0.0, 1.0, 1.0}
\definecolor{dodgerblue}{rgb}{0.12, 0.56, 1.0}
\definecolor{Turquoise}{HTML}{00B4CE}
\definecolor{Periwinkle}{HTML}{7977B8}
\definecolor{Magenta}{HTML}{EC008C}
\definecolor{Thistle}{HTML}{D883B7}
\definecolor{brightblue}{HTML}{0c6efd}
\definecolor{SpringGreen}{HTML}{C6DC67}
\definecolor{NiceOrange}{HTML}{FF6300}
\definecolor{Green}{HTML}{00A64F}

\makeatletter
\def\@listi{\leftmargin\leftmargini
    \parsep 1\p@ \@plus0\p@ \@minus\p@
    \topsep 2\p@   \@plus0\p@ \@minus\p@
    \itemsep1\p@ \@plus0\p@ \@minus\p@}
\let\@listI\@listi\@listi
\makeatother

\renewcommand{\paragraph}[1]{\vspace{0.08in}\noindent\textbf{#1}}

\newcommand{\numparticipants}{16,829\xspace}
\newcommand{\numcountries}{16\xspace}
\newcommand{\numpapers}{151\xspace}
\newcommand{\numfocuspapers}{47\xspace}
\newcommand{\platform}{Facebook\xspace}

\newtcbox{\trendbox}[1][blue!30]{on line,
    colback=#1, colframe=#1, boxsep=0pt, boxrule=0pt, size=small, arc=1mm,right=-1pt,left=-1pt,top=-1pt, bottom=-1pt}

\newcommand{\guidelineFactorSelection}{{G1}}
\newcommand{\guidelineGroupSelection}{{G2}}
\newcommand{\guidelineMethodSelection}{{G3}}
\newcommand{\guidelineResultInterpretation}{{G4}}
\newcommand{\guidelineAuthorPositionality}{{G5}}

\newcommand{\guidelineBox}[1]{\trendbox[gray!30]{\textbf{#1}}}

\newcommand{\trendx}{\trendbox{\textbf{T\#}}\xspace}

\newcommand{\trendGender}{\trendbox{\textbf{T1}}}
\newcommand{\trendAge}{\trendbox{\textbf{T2}}}
\newcommand{\trendAgeSP}{\trendbox{\textbf{T3}}}
\newcommand{\trendEducation}{\trendbox{\textbf{T4}}}
\newcommand{\trendExpertise}{\trendbox{\textbf{T5}}}
\newcommand{\trendGeography}{\trendbox{\textbf{T6}}}

\newcommand{\oppx}{\trendbox[red!30]{\textbf{O\#}}\xspace}
\newcommand{\opportunityGender}{\trendbox[red!30]{\textbf{O1}}}
\newcommand{\opportunityGenderSelfReport}{\trendbox[red!30]{\textbf{O2}}}
\newcommand{\opportunityEducation}{\trendbox[red!30]{\textbf{O3}}}
\newcommand{\opportunityLocation}{\trendbox[red!30]{\textbf{O4}}}
\newcommand{\opportunityGeoLevels}{\trendbox[red!30]{\textbf{O5}}}
\newcommand{\opportunityRace}{\trendbox[red!30]{\textbf{O6}}}
\newcommand{\opportunityIncome}{\trendbox[red!30]{\textbf{O7}}}

\newcommand{\opportunityAuth}{\trendbox[red!30]{\textbf{O8}}}
\newcommand{\opportunityPhish}{\trendbox[red!30]{\textbf{O9}}}
\newcommand{\opportunityAgePhish}{\trendbox[red!30]{\textbf{O10}}}
\newcommand{\opportunitySkill}{\trendbox[red!30]{\textbf{O11}}}
\newcommand{\opportunityTech}{\trendbox[red!30]{\textbf{O12}}}
\newcommand{\opportunityGeoMixed}{\trendbox[red!30]{\textbf{O13}}}
\newcommand{\opportunityBehaviorType}{\trendbox[red!30]{\textbf{O14}}}
\newcommand{\opportunitySelfReport}{\trendbox[red!30]{\textbf{O15}}}
\newcommand{\opportunityDeclare}{\trendbox[red!30]{\textbf{O16}}}
\newcommand{\opportunityBalance}{\trendbox[red!30]{\textbf{O17}}}

\newtcbox{\nameboxturq}[1][Turquoise]{on line,
    colback=#1, colframe=#1, boxsep=0pt, boxrule=0pt, size=small, arc=1mm,right=-1pt,left=-1pt,top=-1pt, bottom=-1pt}
\newcommand{\rachelhong}{\nameboxturq{Rachel Hong}}

\newtcbox{\nameboxthistle}[1][Thistle]{on line,
    colback=#1, colframe=#1, boxsep=0pt, boxrule=0pt, size=small, arc=1mm,right=-1pt,left=-1pt,top=-1pt, bottom=-1pt}
\newcommand{\christinayeung}{\nameboxthistle{Christina Yeung}}

\newtcbox{\nameboxperi}[1][Periwinkle]{on line,
    colback=#1, colframe=#1, boxsep=0pt, boxrule=0pt, size=small, arc=1mm,right=-1pt,left=-1pt,top=-1pt, bottom=-1pt}
\newcommand{\alexandraemichael}{\nameboxperi{Alexandra E. Michael}}

\newtcbox{\nameboxmag}[1][Magenta]{on line,
    colback=#1, colframe=#1, boxsep=0pt, boxrule=0pt, size=small, arc=1mm,right=-1pt,left=-1pt,top=-1pt, bottom=-1pt}
\newcommand{\matthiasfassl}{\nameboxmag{Matthias Fassl}}

\newtcbox{\nameboxblue}[1][Green]{on line,
    colback=#1, colframe=#1, boxsep=0pt, boxrule=0pt, size=small, arc=1mm,right=-1pt,left=-1pt,top=-1pt, bottom=-1pt}
\newcommand{\maximilian}{\nameboxblue{Maximilian Golla}}

\newtcbox{\nameboxsg}[1][SpringGreen]{on line,
    colback=#1, colframe=#1, boxsep=0pt, boxrule=0pt, size=small, arc=1mm,right=-1pt,left=-1pt,top=-1pt, bottom=-1pt}
\newcommand{\aleksei}{\nameboxsg{Aleksei Stafeev}}

\newtcbox{\nameboxdand}[1][NiceOrange]{on line,
    colback=#1, colframe=#1, boxsep=0pt, boxrule=0pt, size=small, arc=1mm,right=-1pt,left=-1pt,top=-1pt, bottom=-1pt}
\newcommand{\alannah}{\nameboxdand{Alannah Oleson}}

\begin{document}

\title{SoK (or SoLK?): On the Quantitative Study of \\ Sociodemographic Factors and Computer Security Behaviors}

\author{
{\rm Miranda Wei}\\
University of Washington
\and
{\rm Jaron Mink}\\
University of Illinois\\ at Urbana-Champaign
\and
{\rm Yael Eiger}\\
University of Washington
\and
{\rm Tadayoshi Kohno}\\
University of Washington
\and
{\rm Elissa M. Redmiles}\\
Georgetown University
\and
{\rm Franziska Roesner}\\
University of Washington
} 

\maketitle

\pagenumbering{gobble}

\begin{abstract}

Researchers are increasingly exploring how gender, culture, and other sociodemographic factors correlate with user computer security and privacy behaviors.
To more holistically understand relationships between these factors and behaviors, we make two contributions.
First, we broadly survey existing scholarship on sociodemographics and secure behavior (\numpapers{} papers) before conducting a focused literature review of \numfocuspapers{} papers 
to synthesize what is currently known and identify open questions for future research.
Second, by incorporating contemporary social and critical theories, we establish guidelines for future studies of sociodemographic factors and security behaviors that address how to overcome common pitfalls.
We present a case study to demonstrate our guidelines in action, at-scale, that conduct a measurement study of the relationships between sociodemographics and de-identified, aggregated log data of security and privacy behaviors among \numparticipants{} users on \platform{} across \numcountries{} countries. 
Through these contributions, we position our work as a systemization of a \textit{lack} of knowledge (SoLK).
Overall, we find contradictory results and vast unknowns about how identity shapes security behavior. 
Through our guidelines and discussion, we chart new directions to more deeply examine how and why sociodemographic factors affect security behaviors.
\end{abstract}


\section{Introduction}
\label{sec:intro}
Sociodemographic factors \dash people's social, cultural, or demographic attributes (e.g., gender, race, socioeconomic status, age, or internet skill) \dash shape their lived experiences, i.e., what happens in their lives, how they are impacted by what happens, and how they make decisions.
Prior works find that sociodemographics \emph{do} impact computer security behaviors, e.g., that women may choose weaker passwords than men~\cite{mazurek_measuring_2013} or that older users choose stronger passwords~\cite{bonneau_science_2012}.
These findings suggest that gender and age influence password selection and thereby computer security, potentially motivating interventions that target the underlying causal mechanisms.

The field of computer security has considered the role of the human for decades, e.g., Saltzer and Schroeder's recognition of the importance of psychological acceptability of security solutions in the 1970s~\cite{saltzer1975protection}, and Whitten and Tygar's foundational 1999 paper catalyzed the formation of the field of \textit{usable security}~\cite{whitten_why_1999}. Focus on the role of sociodemographic factors in computer security behaviors is, however, comparatively new~\cite{redmiles2023critical}. 
Given the potential impact of these factors, it is vital to examine the current state of knowledge with respect to sociodemographics and computer security behaviors.\footnote{For brevity, we use 'security' to consistently refer to security and privacy.} With this understanding, it becomes possible to focus future efforts on addressing knowledge gaps and, ultimately, to help improve computer security for everyone.

Our first two research goals are:
\begin{itemize}
    \item \textbf{Goal 1.} \label{goal:one} Collect and synthesize current knowledge about the quantitative relationship between sociodemographic factors and computer security behaviors.
    \item \textbf{Goal 2.}\label{goal:two} Enumerate existing knowledge gaps about sociodemographics and computer security behaviors.
\end{itemize}

Through a focused literature review of \numfocuspapers papers in selected technical security conferences and a high-level survey of \numpapers papers in the wider literature, we synthesize trends, e.g., that people of different genders may focus on different security behaviors, as well as identify open opportunities for future research. We focus on quantitative studies, a primary method researchers use to measure security behaviors in relation to specific sociodemographic factors, like gender. 
Knowledge gaps exist when pertinent sociodemographic factors are omitted in analyses. 
For example, in our set of \numfocuspapers papers between 1999 and 2023, we find that 38 consider (binary) gender whereas only 3 consider income, 5 consider race, and 9 consider Internet skill; none consider non-binary gender.
We also observe different levels of depth with respect to how sociodemographic factors are analyzed and how differences by factors (if any) are interpreted.

After reviewing the current state of knowledge sociodemographics and behaviors, we identify our third goal:
\begin{itemize}
    \item \textbf{Goal 3.} Formulate guidelines for future research on sociodemographic factors and security behaviors.
\end{itemize}

To demonstrate the use of our guidelines in practice and at-scale, we apply the guidelines to conduct and report the results of a case study.
Our measurement study uses de-identified, aggregated log data from \platform to analyze the relationship between security behaviors on the platform and selected sociodemographic factors.
We confirm several trends observed through our literature review, while adding nuance to others. Finally, we critically consider the knowledge gaps illuminated by our investigation --- particularly, the lack of understanding about \textit{why} sociodemographic factors and security behaviors might be correlated --- and chart directions for future research.


\section{Background and Motivation}
\label{sec:background}
Demographics uses statistics to study trends in human populations~\cite{preston_demography_2001, poston_population_2019}.
Sociodemographics encompass demographic factors as well as social factors defined by formal institutions, e.g., governments~\cite{scott_seeing_2020}, or informal institutions, e.g., sociocultural norms~\cite{saenz_race_2019}.
Conventional studies of sociodemographic factors are positivist, i.e., asserts that knowledge can be empirically measured and there \textit{exists} a correct measurement that scientists can strive for~\cite{ortega_toward_2023}.
In presuming objectivity, conventional demography overlooks the historical and political processes that shaped 
the categories themselves~\cite{horton_critical_1999, saenz_race_2019, sigle_demography_2021}.

\paragraph{Categorization abstracts away richness to allow scientists to focus on selected characteristics.}
As an inherently reductive activity, categorization renders some research more tractable but may not accurately represent lived realities~\cite{bowkers_classification_2000, hoffman_inclusion_2020}.
By assigning people to static, finite groups, those who shift between groups or exist outside those groups, for example, will be systematically misinterpreted~\cite{bivens_gender_2017, keyes_misgendering_2018, scheuerman_we_2020}.
Further, categorization schemes are typically designed by historically and socially privileged groups in ways that can embed power imbalances~\cite{anthro_statement_1998, smedley_race_2005, saini_superior_2019, poston_population_2019}.

\paragraph{\textit{Critical demography}, as an alternative to conventional demography, incorporates the reflexive study of how categories are socially constructed.}
As such, it ``necessitates an open discussion and examination of \textit{power} in society. Specifically, critical demography elucidates how power both affects and is impacted by demographic processes and events''~\cite{horton_critical_1999}.
Thus, critical demography offers a theory-driven paradigm to study how people behave, informed by social and political history~\cite{horton_critical_1999}, towards epistemological diversity and addressing inequity~\cite{ortega_toward_2023}.
Prior work has applied critical demography approaches to deepen knowledge and practice, e.g.,
in computing education~\cite{oleson_demographics_2022}.
We apply a critical demography approach to synthesize prior work on sociodemographic differences 
in security behaviors, but also to map what is not yet known.


\section{Literature Review Methods}
\label{sec:lit-review}

\begin{table*}[!htb]
    \scriptsize
    \caption{Summary of literature review search methods. We used Google Scholar to write custom \textbf{search strings} to identify relevant studies in selected computer science \textbf{venues} and \textbf{databases} as well as in non-CS venues. For each search, we show the \textbf{number of results} and \textbf{number of included} studies satisfying our scoping criteria. \dag We implemented manual keyword searches of PDFs scraped from ndss-symposium.org. *We manually reviewed only the first 3 search result pages. }
    \begin{tabular}{p{2.5cm} p{2.5cm} p{7.5cm} r r} 

    \toprule
    \textbf{Venue} & \textbf{Database} & \textbf{Search String} & \textbf{\# Results} & \textbf{\# Included} \\
    \midrule

    \multicolumn{3}{l}{\textsc{Focus Dataset}: Selected Computer Science Venues} & & \\
    \midrule

    ACM CHI
        & ACM DL
        & {\small In abstract: [\texttt{security} OR \texttt{privacy}] AND [\texttt{behavior} OR \texttt{habits} OR \texttt{practices}] and in body text: [\texttt{gender} OR \texttt{sex} OR \texttt{age} OR \texttt{technical expertise} OR \texttt{education} OR \texttt{race} OR \texttt{culture} OR \texttt{internet skill}]}
        & 213
        & 14 \\ 

    IEEE S\&P
        & IEEE Xplore
        & \textit{same as CHI}
        & 90
        & 2 \\

    USENIX Security
        & USENIX.org
        & {\small [\texttt{security} OR \texttt{privacy}] AND [\texttt{behaviors} OR \texttt{habits} OR \texttt{practices}]}
        & 41
        & 2 \\

    SOUPS
        & ACM DL 
        & \textit{same as CHI} \newline 
        & 77 
        & 17 \\ 

    ACM CCS
        & ACM DL
        & \textit{same as CHI}
        & 508
        & 7 \\

    ACM CSCW
        & ACM DL
        & \textit{same as CHI}
        & 62
        & 4 \\

    NDSS
        & ndss-symposium.org\dag        & \textit{same as CHI}
        & 62
        & 1 \\

\midrule
    \multicolumn{3}{l}{\textsc{Full Dataset:} Beyond Selected Computer Science Venues} & \\
    \midrule

    \textit{Various}
        & Google Scholar
        & {\small [\texttt{cross cultural} OR \texttt{large scale} OR \texttt{demographic}] AND [\texttt{behaviors}] AND [\texttt{security} OR \texttt{privacy}]} 
        & 270K+*
        & 97 \\
        
    \textit{Various}
        & Google Scholar
        & {\small [\texttt{password} OR \texttt{authentication} OR \texttt{update software} OR \texttt{secure drop} OR \texttt{phishing emails} OR \texttt{encryption} OR \texttt{WiFi} OR \texttt{anti-virus} OR \texttt{HTTP SSL warnings} OR \texttt{tracker blockers} OR \texttt{information disclosure} OR \texttt{self disclosure} OR \texttt{IoT} OR \texttt{VPN}] AND [\texttt{behaviors}]}. 
        & 5.8M+*
        & 11 \\

    \bottomrule
    \end{tabular}
    \label{tab:lit-review}
\end{table*}
What is currently known about how sociodemographics affect behavior, and what gaps remain?
To scope to studies of users' actual security or privacy behaviors, we excluded studies of intended behavior, concerns, knowledge, or attitudes.
As we were also interested in quantitative studies, we only included works that compared behavior between sociodemographic groups, i.e.,
we excluded work that investigated only one group within a sociodemographic factor.

\subsection{Identifying Relevant Work}
To identify potentially relevant studies, we defined unique search queries for selected conferences (see Section~\ref{ss:methods-datasets}) and used the advanced search features of the ACM DL, IEEE Xplore, and the USENIX databases to search full-length research articles (see Table~\ref{tab:lit-review}).
Since these databases do not contain NDSS papers, we also obtained an NDSS paper archive scraped by other researchers.
We wrote a \texttt{pypdf}~\cite{pypdf} script to extract and search text directly from the PDFs using the search strings shown in Table~\ref{tab:lit-review}.
Two researchers independently reviewed paper titles and abstracts of search results to apply the scoping criteria described above and iteratively resolved disagreements to select the final dataset.

Relevant studies are also published in venues beyond computer science conferences; we used Google Scholar to find popular studies from any venue, including journals of business, information science, social science, or grey literature. We then defined two sets of keyword searches (see Table~\ref{tab:lit-review}), which yielded over 6 million results, so one researcher reviewed the first 3 pages of search results.
Finally, we followed citations from papers in our dataset that referred to relevant work, adding 20 studies not identified through search strings. 
We set no explicit time boundaries for our dataset.

\subsection{Full and Focus Datasets}
\label{ss:methods-datasets}
Our final ``full'' dataset consisted of \numpapers{} works. Most papers were published in academic venues such as conferences or journals, but we also included 4 theses, 3 Pew Research studies, and 1 arXiv paper.
The full dataset reflects a growing interest in security behaviors with respect to sociodemographic factors across venues and academic disciplines.
Much of the dataset (76 papers) was in information science or social science domains and spanned a wide range of venues, from computing (e.g., \textit{Computers in Human Behavior}) to communications and media (e.g., \textit{New Media \& Society}) to marketing and business (e.g., \textit{Journal of Interactive Marketing} and \textit{Journal of Management Information Systems}) to social sciences (e.g., SSRN). 
Another 20 papers were in computer science publications. The distribution of venues is long-tailed since we had 70 papers each from unique domains.
We defined a ``focus'' set of \numfocuspapers{} papers by identifying seven conferences most likely to include papers of interest: four computer security conferences (IEEE S\&P, USENIX Security, CCS, NDSS), two HCI conferences with a tradition of including security and privacy (ACM CHI, ACM CSCW), and one conference at the intersection of HCI and security (SOUPS). 

\subsection{Qualitative Analysis}
\label{ss:qualitative}
We qualitatively analyzed papers in our ``full'' set by coding behaviors studied (dependent variables) and sociodemographic factors considered (independent variables).
For further analysis on our ``focus'' set, we also coded whether a significant relationship was (or was not) found\footnote{Because studies sometimes studied multiple sociodemographic factors for a given behavior, we coded each relationship separately.} as well as where the study was conducted, the research methods used, and any research sample limitations. 

We created our codebooks via a series of iterative coding sessions between two coders.
First, a primary coder prepared an initial codebook by inductively coding all papers. A second coder then independently coded a subset of the papers using the same codebook. The two coders then met to resolve inconsistencies and, if necessary, clarify and adjust the codebook. If adjustments were made, the primary coder then recoded the rest of the dataset. This process was repeated until the codebook no longer changed.
We also verified our resulting codebook with a prior work's codebook on security behaviors~\cite{redmiles_comprehensive_2020}.
Table~\ref{tab:counts-behaviors} presents the behaviors codebook and paper counts; codebooks for other topics are presented in Appendix~\ref{sec:appendix-codebooks}.

\subsection{Positionality}
\label{ss:positionality}
The authors' particular social, cultural, political, and historical context influence the way we discuss sociodemographic factors in this work.
As researchers who have predominately lived and worked in the U.S., have English as a first language, and have had opportunities to pursue or achieve academic degrees in computer science, our perspective is limited by the privileges these experiences afford, relative to different experiences. Our motivation for this work is also shaped by experiences of marginalization by gender, culture, race, and age. As researchers with substantial experience studying human factors in security, we aim to improve the security of all people, not only those who have been historically prioritized in security research (i.e., users who are predominantly men, white, wealthy, highly educated, and live in the U.S.).
We seek to raise the voices of those at the margins, in alignment with standpoint theory's premise that non-dominant social groups contribute critical knowledge towards scholarship and action towards justice~\cite{collins_1998_fighting}.

\subsection{Limitations}
\label{ss:limitations}

We sought as exhaustive a list of papers as possible to study sociodemographic factors and security behaviors, but we likely missed some relevant papers. 
During paper collection, we ultimately included less than 25\% of our search results because many papers use sociodemographic keywords 
without satisfying our criteria, i.e., they do not quantitatively compare groups within a factor.
Relevant work is also published outside the seven venues of our focus papers, but we believe our methods captured a set of papers large enough for meaningful analysis and discussion. 

Our goal was to focus on sociodemographic factors related to security behaviors. 
As such, we scope to papers that measure security behaviors directly, e.g., through observational data, log data, and self-reports about actual behaviors.
Future work may seek to focus on the substantial literature on additional topics, such as attitudes, opinions, or perceptions.

Since our goal was to conduct a formative literature review of security behaviors with respect to sociodemographic factors, we did not attempt to evaluate the ``validity'' of any paper.
The replication crisis in psychology~\cite{shrout_psychology_2018} reminds us that robust quantitative findings must be repeatedly tested and confirmed, which we leave to future work.


\section{Literature Review Results}
\label{sec:lit-results}

We first survey the locations, methods, and behaviors investigated in the full dataset of \numpapers{} papers (Section~\ref{ss:lit-results-full}) and then explore methodological considerations in our \numfocuspapers{} focus papers (Section~\ref{ss:lit-results-focus}).
Next, we synthesize results about eight sociodemographic factors---gender, age, education, technical expertise, Internet skill, geography, race, and income---from the focus papers (Sections~\ref{ss:results-gender}-\ref{ss:results-income}).
Table~\ref{tab:crosstabs} provides an overview of sociodemographic factors studied with respect to security behaviors and whether differences among groups were found.

Throughout this section, we enumerate trends as \trendx and opportunities for future work as \oppx to facilitate later comparison with our measurement study results. 
We present a summary of all trends and opportunities in Table~\ref{tab:trends-opportunities}.

\subsection{Survey of Full Dataset}
\label{ss:lit-results-full}
\begin{figure}
    \centering
    \resizebox{\columnwidth}{!}{
        \includegraphics[trim={0 0.25cm 0 0cm},clip]{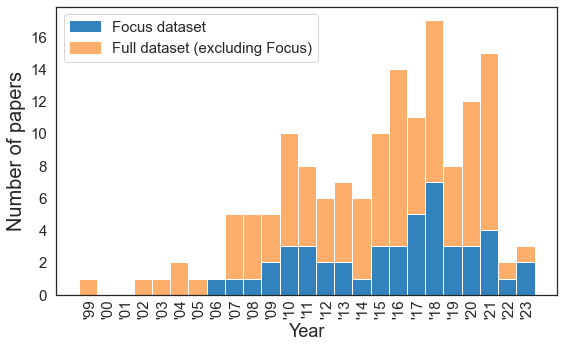} }
    \caption{Number of papers in focus and full datasets investigating sociodemographics and security behaviors over time.
    }
    \label{fig:studies-over-time}
\end{figure}

Since 1999, the publication year of the first study in our dataset, papers on sociodemographics and security have steadily increased (see Figure~\ref{fig:studies-over-time}). 

\paragraph{Most studies are conducted in the U.S.\ or Western Europe \opportunityLocation{}.}
A plurality of studies in the full dataset (N=\numpapers{}) were conducted in the U.S.\ or Western Europe (60), followed by Asia (20), Africa and the Middle East (9), Australia (6), and South America (3).\footnote{Counts do not sum to \numpapers{} due to studies conducted in multiple locations.}
Studies in the focus dataset (N=\numfocuspapers{}) show the same trend: U.S.\ or Western Europe (33), Asia (5), Africa and the Middle East (3), Australia (3), South America (2).

\begin{table}[tb]
    \footnotesize
    \caption{The number of security behaviors studied by papers in our \textbf{full} (N=\numpapers) and \textbf{focus} (N=\numfocuspapers) dataset. Counts do not sum to \numpapers{} or \numfocuspapers{} due to papers that study multiple behaviors.}
    \vspace{0.5em}
    \begin{tabularx}{\columnwidth}{p{22em}XX}
    \toprule
        \textbf{Security Behavior} & \textbf{Full} & \textbf{Focus} \\
    \midrule
        \textbf{Network and Web}, e.g., VPNs, use private browsing, use anti-tracker tools
            & 44
            & 9 \\
        \textbf{Phishing and Spam}, e.g., phishing susceptibility
            & 38
            & 9 \\
        \textbf{Social Sharing}, e.g., disclosing info. on social media, changing privacy settings
            & 35
            & 11 \\
        \textbf{Authentication and Accounts}, e.g., password creation or reuse, 2FA
            & 23
            & 12 \\
        \textbf{Device}, e.g., anti-virus software, mobile lock screens
            & 21
            & 8 \\
        \textbf{Composite}: combined security behaviors from above that cannot be disaggregated
            & 14
            & 5 \\
    \bottomrule
    \end{tabularx}
    \label{tab:counts-behaviors}
\end{table}
\paragraph{Studies reflect a significant interest in network and web security behaviors as well as phishing and spam, but comparatively less in other behavior types \opportunityBehaviorType.}
A complete breakdown of behaviors studied by papers in our full and focus datasets is presented in Table~\ref{tab:counts-behaviors}.
The largest portion of our full dataset investigated network and web-related behaviors (44 papers), such as using VPNs or avoiding public WiFi, or using private browsing and other anti-tracking tools or practices.
The second largest portion investigated phishing or spam susceptibility, a much narrower set of behaviors than network and web yet studied by nearly as many studies (38 papers).
Less commonly studied were social sharing behaviors, e.g., disclosing information or changing privacy settings on social media (35), authentication and account behaviors (23), and device-related behaviors (21).
Finally, 14 papers studied composite security behaviors, or combinations of the above types of behaviors, with regression analyses that we could not disaggregate.

\paragraph{Security behaviors are primarily investigated through self-report methods \opportunitySelfReport.}
For our full dataset, the vast majority, i.e., 109 studies, were based on self-reports of security behavior.
Less common were the 32 studies that used experimental measurement methods, e.g., in-lab or online experiments, or the 15 that used observational methods, e.g., log data or installed software on user devices.
For the focus dataset, 32 were self-reports, 12 used experimental measurement methods, and 8 used observational methods.
The significant preference for self-report methods likely reflects the relative convenience of collecting data from participants simply by asking them, but self-reporting may not be wholly accurate given participant biases, e.g., social desirability biases~\cite{redmiles2017summary}.
Future work should confirm these results with more ecologically valid methods.

\subsection{Methods Considerations in Focus Dataset}
\label{ss:lit-results-focus}

For the focus dataset, we were interested in how prior work ensured that the research was robustly designed and conducted for the sociodemographics and security behaviors of interest.

\paragraph{A sizeable proportion of investigations did not mention sociodemographic factors until the results section of the papers \opportunityDeclare.}
We analyzed focus dataset papers to determine if they motivated their study of sociodemographics in the research questions or introductory section (i.e., considered) or whether sociodemographic factors were mentioned only in the results section (i.e., post hoc). 
Because clearly defining independent variables, e.g., gender, and their levels or conditions, e.g., woman/man/nonbinary,  is essential to control for confounding factors between conditions~\cite{purchase2012experimental}, not considering the role of sociodemographics until the results may imply an incomplete methodology.
Further, consistently reporting the variables and relationships of interest before beginning statistical testing is important to avoid cherry-picking non-null results~\cite{mcclure_identifying_2020}. 
We found 74 instances where factors were considered in advance and 45 instances where sociodemographic factors were studied post hoc.\footnote{Papers would have multiple instances if they studied multiple factors; thus, counts do not sum to \numfocuspapers{}.}
Studying sociodemographic relationships to security behavior is not warranted merely because relevant data was collected; therefore, confirming the results of papers that conducted a post hoc study of sociodemographics is an opportunity for future research to ensure that the findings were not spurious.

\paragraph{The majority of papers had limited samples, i.e., samples not representative of broader populations or balanced among groups \opportunityBalance.}
To assess the generalizability of studies in our focus dataset, we coded whether papers were representative (i.e., sample attributes matched broader population attributes), or balanced (i.e., equally across factor groups) for analysis.
We found only 7 instances of factors being balanced and 18 representative; the vast majority (96) were limited,\footnote{Counts do not sum to \numfocuspapers{} because of papers studying multiple factors.}
i.e., not controlled in any way (snowball or convenience samples), or had no description of recruiting considerations. Future work should expand on addressing limited samples.

\subsection{Gender}
\label{ss:results-gender}

Gender is used in varying contexts and includes a person's gender, but also how gender is constructed in a societal context, i.e., referring to socially established gender roles~\cite{keyes_you_2021}.
It is often conflated with sex, i.e., bodily attributes, though these are distinct concepts~\cite{keyes_you_2021}. 
Gender has received the most attention in security research (38 of \numfocuspapers focus papers) relative to other sociodemographic factors. 

\paragraph{Existing research primarily uses self-report methods, which could be biased by gendered differences in self-reporting \opportunityGenderSelfReport.}
Prior work finds gender stereotypes that men are overconfident when it comes to security and privacy~\cite{wei_stereotypes_23}.
As described above, sociodemographic differences in behavior tend to be investigated through self-report methods, a trend that holds for gender specifically:
of 28 papers that found gender differences, 21 used self-report methods.
Thus, gender differences could be because men are more likely to self-report behaviors than women, regardless of true adoption rates.

\paragraph{Prior work suggests men may focus more on technical security behaviors, while women may focus more on privacy behaviors \trendGender.} 
One set of papers found that men were more likely to take certain protective actions related to network and web security, e.g., use tracker blockers~\cite{mathur_characterizing_2018}, take protective actions against trackers~\cite{coopamootoo_invaded_2022}, and use private browsing~\cite{habib_away_2018}, although no differences were found for heeding SSL warnings~\cite{sotirakopoulos_challenges_2011}.
Two papers studied composite security behaviors, finding that men were more likely to adopt predefined sets of security and privacy protective practices~\cite{zou_examining_2020, wash_too_2015} (although no differences were found in ``triggers'' prompting security behaviors~\cite{das_typology_2019}).
Finally, one paper investigated how the sources for security advice differ between genders~\cite{redmiles_secure_2016}, finding that men were more likely than women to source advice from service providers.
Taken together, these findings suggest gender differences in ``technical'' security behaviors, though it is unclear whether these differences result from self-reporting biases, prior computing experience, attitudes towards computers~\cite{williams_gender_1993}, or something else.

Another set of papers found that women were more likely to engage in security and privacy behaviors on social media and personal devices.
Women were more likely to have private profiles on Facebook or Myspace~\cite{gilbert_network_2008, stutzman_friends_2010}, post non-publicly on Facebook~\cite{fiesler_what_2017} and Snapchat~\cite{habib_snapchat_2019}, and avoid actions that expose online profiles they viewed~\cite{hoyle_viewing_2017}.
Teen girls were found to be more likely than teen boys to adopt risk-coping behaviors (e.g., deleting posts, untagging photos, faking personal information) as well as seek privacy advice~\cite{jia_risk_2015}.
Prior work also found differences in disclosure content: men were more likely to disclose COVID-19 distress in May 2020 than women~\cite{zhang_distress_2021}, but generally women were more likely to share memes portraying subjects positively~\cite{hasan_your_2021}.
Though prior work found inconclusive evidence about gender differences in device behaviors --- adopting lock screens~\cite{harbach_keep_2016, van_bruggen_modifying_2013}, Android updating~\cite{mathur_impact_2017} --- in others, women were found to be more likely to use webcam covers~\cite{machuletz_webcam_2018} and women 18-23 were more likely than men or people of other age groups to deny Android permission dialogs~\cite{bonne_exploring_2017}.
Taken together, these findings align with prior work (outside our literature review)~\cite{sambasivan_privacy_2018, madden2012privacy,karusalaPatriarchy2019, tifferet2019gender,redmiles_net_2018} indicating that women focus on information protection and engage in privacy-preserving self-censorship.

\paragraph{Results were mixed on gender differences with respect to authentication and susceptibility to phishing and spam.}
Research on authentication behaviors is mixed: two papers found that men's passwords were stronger against offline attacks~\cite{bonneau_science_2012,mazurek_measuring_2013}, but men aged 46-49 were more likely to share passwords than others~\cite{kaye_self-reported_2011}; another paper did not find sharing differences by gender~\cite{park_share_2018}. Further, researchers found that women were more likely to reuse passwords with slight modifications~\cite{shay_encountering_2010}, less likely to remember graphical passwords~\cite{chiasson_graphical_2009}, and less likely to enable 2FA in response to experimental prompts~\cite{golla_driving_2021}, but other researchers found no difference in password reuse~\cite{pearman_people_2019} or in whether they change password managers~\cite{munyendo_managers_2023}.
Future work should investigate whether gender differences in authentication behaviors are due to methodological differences, context, or other reasons \opportunityAuth.

Two papers found that women were more likely to click on phishing and spam~\cite{redmiles_examining_2018, sheng_who_2010}, although three other papers did not find significant gender differences in this regard~\cite{dhamija_why_2006, kumaraguru_school_2009, sheng_anti-phishing_2007}, and one found that women were less likely to visit malicious URLs than men~\cite{sharif_malicious_2018}.
Papers not finding gender differences were published in 2006, 2007, and 2009, while papers finding differences were published in 2010 and 2018. 
One explanation could be that phishing and spam increasingly targeted women in 2010, and people of different genders now receive different types of phishing and spam~\cite{redmiles_examining_2018} \opportunityPhish.

\paragraph{Existing research primarily investigates binary (assumed cisgender) individuals, excluding non-binary and transgender people \opportunityGender.}
Most papers mention only women and men, and few papers conduct statistical testing with non-binary individuals, often opting to filter them out during data processing.
Non-binary people constitute a far smaller proportion of study participants, posing a challenge for parametric statistical testing that could be resolved with use of nonparametric tests or different study designs. 
Further, almost no papers discuss transgender individuals, while 
other work conflates gender and sex by referring to participants as female and male when discussing gender, against best practices~\cite{scheuerman_guidelines_2020}.
Research should distinguish between cisgender and transgender people only when relevant, but given that transgender people experience significant harm~\cite{vawnet_violence} and erasure~\cite{suarez_resisting_2018}, omitting this aspect of gender may reflect cisnormativity.\footnote{\textit{Cisnormativity} is the assumption that everyone is or should be cisgender.}

\subsection{Age}
\label{ss:results-age}
Age granularity in the security literature varies from a single year to multiple decades and can be modeled as a numeric or categorical variable.
Age is the second most studied sociodemographic factor (in 30 of \numfocuspapers focus papers).

\paragraph{Prior work suggests that age may have been correlated with differences in password behaviors in the past, but is no longer \trendAge.}
Two papers published after 2017 found
no significant differences by age in switching password managers~\cite{munyendo_managers_2023} or 
password reuse~\cite{pearman_lets_2017} (also found by a 2010 paper~\cite{shay_encountering_2010}).
Supporting a theory of change in the past decade, a 2012 paper found that older users chose stronger passwords~\cite{bonneau_science_2012} but a study from the following year did not find such correlations~\cite{mazurek_measuring_2013}; similarly a 2011 paper found older users were more likely to share passwords~\cite{kaye_self-reported_2011}, but evidence from a study seven years later did not support this finding~\cite{park_share_2018},

\paragraph{Older users may behave more securely, while younger users focus on privacy \trendAgeSP.}
When studying a combined set of internet \textit{security} behaviors, prior work found that older adults behaved more securely~\cite{wash_too_2015, zou_examining_2020}, while younger users were more likely to adopt a combined set of \textit{privacy} practices~\cite{zou_examining_2020}, e.g., use private browsing~\cite{habib_away_2018}, use tracker blockers~\cite{mathur_characterizing_2018}, and have Android lock screens~\cite{harbach_keep_2016} (though older users in Singapore may differ as they were more likely to adopt privacy practices~\cite{cho_networked_2016}).
The distinction between security and privacy behaviors may be partially explained by the finding that people 60+ were more likely to learn from automatic requirements or service providers than younger people~\cite{redmiles_secure_2016}:
formal sources may emphasize security as prevention of universal harm but privacy as a personal choice.
Thus, older users were found to be more likely to enable 2FA in response to prompts~\cite{golla_driving_2021} and deny Android permissions dialogs~\cite{bonne_exploring_2017} as well as be more likely to be prompted by social triggers to behave securely~\cite{das_typology_2019}. 
However, differences by age were not found in responses to SSL warnings~\cite{sotirakopoulos_challenges_2011} or Android auto-updating~\cite{mathur_characterizing_2018}.

With respect to online sharing, older users were more likely to post publicly than younger users~\cite{fiesler_what_2017} though less likely to specifically disclose distress during the COVID-19 pandemic~\cite{zhang_distress_2021} or share a security news event~\cite{das_breaking_2018}.

\paragraph{Similar to mixed results for phishing susceptibility by gender, prior work presents inconclusive findings about the relationship between age and phishing susceptibility \opportunityAgePhish.}
Three papers found that younger participants were more susceptible to phishing~\cite{sheng_who_2010, kumaraguru_school_2009, hasegawa_why_2021}, 
while two found no correlations by age~\cite{dhamija_why_2006, sheng_anti-phishing_2007}.

\begin{table*}[ht]
    \scriptsize
    \caption{Relationships between sociodemographic factors and security behaviors for papers in our focus dataset. For each sociodemographic factor (rows) and category of security behaviors (columns), we show X / Y, where X is the number of papers that found differences by factor for behavior, and Y is the total number of papers studying that factor and behavior. Summing counts do not sum to totals due to papers that study multiple factors or behaviors. Auth. = Authentication, Tech. Exp. = Technical Expertise, Composite = multiple behaviors studied together.} \label{tab:crosstabs}
    \begin{tabularx}{\textwidth}{XXXXXXXr}
    \toprule
                        & \textbf{Accounts \newline \& Auth.}
                        & \textbf{Device}
                        & \textbf{Network \newline \& Web}
                        & \textbf{Phishing \newline \& Spam}
                        & \textbf{Social Media \newline \& Sharing}
                        & \textbf{Composite} 
                        & \textsc{\textbf{Total}} \\
    \midrule
    Gender              & 6 / 9       & 4 / 6  & 3 / 4       & 3 / 6         & 10 / 10  & 3 / 4 & 28 / 38  \\
    Age                 & 3 / 9       & 2 / 3  & 2 / 3       & 3 / 5         & 5 / 6   & 4 / 4 & 19 / 30  \\
    Education           & - / 4       & - / 2  & 1 / 2       & - / 2         & - / 1   & 3 / 3 & 4 / 14  \\
    Tech. Exp.          & 3 / 6       & 2 / 3  & 4 / 4       & 4 / 5         & 0 / 0   & 1 / 1 & 7 / 12  \\
    Geography           & 1 / 1       & 1 / 1  & 2 / 2       & 2 / 2         & 2 / 2   & 2 / 2 & 10 / 10  \\
    Internet Skill      & 2 / 2       & - / -  & 1 / 1       & 2 / 4         & 1 / 1   & 1 / 1 & 7 / 9   \\
    Race                & - / -       & - / -  & - / -       & - / -         & 2 / 3   & 2 / 2 & 4 / 5   \\
    Income              & - / 1       & - / -  & - / -       & - / -         & - / -   & 2 / 2 & 2 / 3   \\
    \midrule
    \textsc{Total}      & 12           & 8      & 9            & 10           & 11      & 5     & \numfocuspapers  \\
    \bottomrule
    \end{tabularx}
\end{table*}

\begin{table*}[th]
    \scriptsize
    \caption{A summary of trends and opportunities for future research from our literature review. }
    \vspace{0.5em}
    \begin{tabularx}{\textwidth}{cXcX}
    \toprule
        \textbf{T\#} & \textbf{Trends in Findings} & & \\
    \midrule

        \trendGender &
            \multicolumn{3}{l}{Women seem to focus more on information protection, while men seem to focus more on technical security.} \\
        \trendAge &
            \multicolumn{3}{l}{Older users may have had different password behaviors in the past, but no longer.} \\ 
        \trendAgeSP &
            \multicolumn{3}{l}{Older users seem to exhibit more security-related behaviors while younger users focus more on privacy.} \\
        \trendEducation &
            \multicolumn{3}{l}{Education does not seem to be correlated with secure behavior.} \\
        \trendExpertise &
            \multicolumn{3}{l}{Users with more tech expertise/use seem more likely to adopt technical security tools and take protective actions.} \\
        \trendGeography & 
            \multicolumn{3}{l}{Geography seems to be strongly correlated with differences in security behaviors.} \\
            
    \midrule
        \textbf{O\#} & \textbf{Opportunities for Future Research} & & \\
    \midrule  
    \multicolumn{4}{l}{\textbf{Who is being studied: Lack of Focus around Specific Groups}}  \\ 
    \midrule

        \opportunityGender & 
            Lack of research on non-binary and transgender people's security behaviors. &
            \opportunityGeoLevels{} & 
            Lack of research on geographical differences beyond granularity of countries. \\
        \opportunityGenderSelfReport & 
            Lack of research on gender differences in self-reported behaviors. &
            \opportunityRace & 
            Lack of research on race and security behaviors. \\
        \opportunityEducation & 
            Lack of research on education at levels besides (post-)secondary. &
            \opportunityIncome & 
            Lack of research on income and security behaviors. \\ 
        \opportunityLocation &
            \multicolumn{3}{l}{Majority of papers conducted in U.S.\ and Western contexts; relative lack of research in other locations.} \\

\midrule
\multicolumn{4}{l}{\textbf{What was found: Contradictory or Unclear Results}} \\
\midrule
        \opportunityAuth & 
            Mixed results on authentication behavior $\sim$ gender. &
            \opportunitySkill & 
            Mixed results on phishing susceptibility $\sim$ internet skill. \\ 
        \opportunityPhish & 
            Mixed results on phishing and spam susceptibility $\sim$ gender. &
            \opportunityTech & 
            Mixed results on password behaviors $\sim$ technical expertise.\\
        \opportunityAgePhish & 
            Mixed results on phishing susceptibility $\sim$ age. &
            \opportunityGeoMixed{} &
            Unclear patterns of geographical influence on security behaviors. \\ 
\midrule   
\multicolumn{4}{l}{\textbf{How: Methodological Issues}} \\
\midrule

        \opportunityBehaviorType &
            \multicolumn{3}{l}{Significant interest in network/web behaviors and phishing/spam, but less on other behaviors.} \\
        \opportunitySelfReport &
            \multicolumn{3}{l}{Security behaviors are primarily investigated through self-report methods.} \\
        \opportunityDeclare &
            \multicolumn{3}{l}{Many papers did not declare an interest in sociodemographic factors in the motivation of the work.} \\
        \opportunityBalance &
            \multicolumn{3}{l}{Most papers had limited sample generalizability.} \\

    \bottomrule
    \end{tabularx}
    \label{tab:trends-opportunities}
\end{table*}

\subsection{Education}
\label{ss:results-education}
Formal education imparts knowledge and skills to students and increasingly includes information about computing.
Educational systems and institutions vary greatly, including nationally and internationally, but can be broadly grouped into primary, lower and upper secondary, and tertiary (also called higher ed)~\cite{unesco_isced_2012}.

\paragraph{Education does not seem to be correlated with secure behavior \trendEducation.}
Of the 14 papers that investigated relationships between education and security behaviors, four found significant correlations:
more educated users were more likely to delete cookies and history~\cite{chen_exploring_2014} as well as adopt a composite of 30 security, privacy, and ID theft practices~\cite{zou_examining_2020}.
However, studying a composite of four behaviors to combat viruses and hackers, Wash et al. find that compared to those with a high school diploma, those who did not complete high school were more likely to adopt security behaviors~\cite{wash_too_2015}. Similarly, compared to those who held a BA, those who did not hold a BA were more likely to report learning security advice from automatic software updates~\cite{redmiles_secure_2016}.

On the other hand, 10 papers do not find significant correlations between education and account sharing~\cite{park_share_2018}, password strength~\cite{wimberly_fingerprint_2011}, password reuse~\cite{pearman_lets_2017}, switching of password managers~\cite{munyendo_managers_2023}, Android auto-updating~\cite{mathur_impact_2017}, webcam cover use~\cite{machuletz_webcam_2018}, 
public sharing behaviors on Snapchat~\cite{habib_snapchat_2019},
SSL warning behaviors~\cite{sotirakopoulos_challenges_2011}, or phishing susceptibility~\cite{dhamija_why_2006, kumaraguru_school_2009}.

\paragraph{Differences in correlations between education and security behavior are not well understood.} 
There may exist several reasons for these disparate results.
First, while prior work notes that those with lower educations are  more concerned about being the victim of a computer scam, losing financial information, and being the target of harassment~\cite{madden2017privacy}, it is unclear how the varying computer knowledge held by those with different educational backgrounds affects the ability to employ secure behaviors.
Education does not necessarily include computing or security education; indeed, prior work found that while people with less education rely on less authoritative sources of security advice, they report fewer negative incidents, perhaps suggesting that formal advice sources --- including formal educational environments --- fail to provide effective security education~\cite{redmiles_digital_2017}.
Further, a bachelor's degree education varies significantly by institution, such that high-level education categories reduce critical nuances.
Education may not result in a linear increase in security behavior but may vary by context.
Future work should investigate the relationships between education and security behaviors to better understand the underlying causal mechanisms at play.

\paragraph{There is a lack of research on students at levels besides secondary or post-secondary \opportunityEducation.}
Of 14 papers studying education and security behaviors, four conducted studies in U.S. and Canadian universities (e.g., university students, staff, faculty) and another seven conducted studies with U.S.\ recruitment/crowdworker populations, which are more likely to have attended or completed college than the average~\cite{pew_courdsourcing_2016}.
Only the remaining three papers were not conducted in the U.S.\ or Canada, revealing a striking over-representation of Western university-affiliated users in education-related results.
Future work should consider a wider range of educational levels, in different or outside of educational systems.

\subsection{Technical Expertise, Use, and Skill}
\label{ss:results-tech-internet}
Aside from general education, users have varying levels of experience with technology (i.e., technical expertise) or the internet specifically (i.e., internet skill).

\paragraph{Users with more technical expertise may use more technical security tools and take more protective actions \trendExpertise.}
People with greater technical expertise were found to be more likely to use private browsing~\cite{habib_away_2018}, identify security threats~\cite{onarlioglu_insights_2012}, and cite school (as opposed to required sources or device prompts) as a source of security advice~\cite{redmiles_secure_2016}.
Those with computer and mobile skills were more likely to take defensive security measures~\cite{chen_exploring_2014}.
Greater technical expertise was also associated with higher adoption of multiple security practices~\cite{ion_no_2015, busse_replication_2019}, although no correlation was found between technical expertise and webcam cover use~\cite{machuletz_webcam_2018}.

Relatedly, internet use may suggest higher adoption of security practices, e.g., users who logged in from multiple locations chose stronger passwords~\cite{bonneau_science_2012}, and users more active on Facebook were more likely to enable 2FA in response to prompts~\cite{golla_driving_2021}.
Prior work also demonstrates that people with more internet skill cite different sources of advice~\cite{redmiles_secure_2016}, which may contribute to these behavioral differences. 

\paragraph{We observe an inconclusive relationship between technical expertise and password-related behaviors \opportunityTech.}
Unlike other security behaviors, technical expertise did not have a clear relationship with password choices. Two papers studied people affiliated with universities, one finding that participants in the computer science department chose stronger passwords than those in business~\cite{mazurek_measuring_2013}, but the other found no conclusive evidence that technical expertise (including departmental affiliation) was correlated with stronger passwords~\cite{wimberly_fingerprint_2011}.
Further, two papers found no correlation between technical expertise and password reuse~\cite{shay_encountering_2010} or password manager switching~\cite{munyendo_managers_2023}.

\paragraph{There is an inconclusive relationship between internet skill and phishing susceptibility \opportunitySkill.}
Two papers found that greater internet skill or knowledge about phishing correlated with less phishing or spam susceptibility~\cite{redmiles_examining_2018, sheng_who_2010}, while three others did not find correlations between internet skills or phishing susceptibility~\cite{dhamija_why_2006, sheng_anti-phishing_2007,hasegawa_why_2021}. 
A potential explanation comes from outside the literature review: prior work suggests activity level on a platform (which is itself weakly correlated to internet skill) may have more explanatory power than the coarser measure of internet skill~\cite{redmiles_examining_2018}.

\subsection{Geography and Race}
\label{ss:results-geography-race}
Geography is a proxy and umbrella term for a range of sociodemographic factors, including nationality, language, population density, political history, culture, internet penetration, freedom of speech, and more. 
Geography also shifts over time since politics and culture reshape the societies living between socially constructed boundaries.

Race refers to groups of people who share cultural, social, and physical similarities.
It has been shaped through historical narratives of identity to be a tool of power, particularly for discrimination and the justification of colonialism~\cite{anthro_statement_1998, smedley_race_2005, saini_superior_2019}.
Though racialized science continues to advance myths of biological differences between races, race is a powerful determinant of the privileges that an individual has access to, e.g., education, wealth, health.

\paragraph{All papers in our focus dataset studying geographic factors with respect to security behaviors were significantly correlated with behaviors \trendGeography{}, but effects lacked clear cross-cultural patterns  \opportunityGeoMixed{}.}
While the ten papers investigating correlations between geography and security behaviors find differences in many types of behaviors, these results are often sparsely populated, and it is not clear why these patterns appear or how they do, or do not, generalize to other geographical regions.
German and French participants were found to be twice as likely to take protective actions against tracking than those in the UK~\cite{coopamootoo_invaded_2022}.
Compared to U.S. and U.K. users, German internet users tended to adopt more advanced, active privacy methods, such as proxies, Tor, and providing false information~\cite{coopamootoo_pet_2020}.
U.S. users were more likely to take security-protective actions because of proactive triggers, whereas people in India were more likely to act in response to social triggers~\cite{das_typology_2019}.
Password strength varied by primary language spoken: passwords chosen by Indonesian-speaking users were found to be the weakest; German- and Korean-speaking users tended to choose relatively strong passwords~\cite{bonneau_science_2012}.
Android lock screen usage also varied by country, e.g., 76.4\% in the U.K. compared to 50.4\% in Italy~\cite{harbach_keep_2016}.

Phishing and spam susceptibility also differed by geographic location. Users who live in countries that have more spam are less likely to click it~\cite{redmiles_examining_2018}. 
South Koreans were more likely to fall for phishing attacks in Korean than English, while Japanese participants were more likely to fall for phishing in English than Japanese~\cite{hasegawa_why_2021}.
On social media, rural U.S. users were more likely to set profiles to private than urban U.S. users~\cite{gilbert_network_2008}, and Saudi women were more likely to block people on WhatsApp than Indian women~\cite{dev_lessons_2020}.  Compared to U.S. users, U.K. users were less likely to dismiss cookie banners but more likely to not read consent text~\cite{bouma-sims_cookie_2023}.

\paragraph{Few papers discuss geographical factors beyond the granularity of a country \opportunityGeoLevels{}.}
Geographical factors describe a wide range of sociodemographic variance beyond national identity; however, the majority of papers focus only on these differences.
Of ten papers, eight segregate geographical differences by nationalities~\cite{bouma-sims_cookie_2023, coopamootoo_invaded_2022,das_typology_2019, dev_lessons_2020, redmiles_examining_2018, harbach_keep_2016,hasegawa_why_2021}, while only two discuss variations by language spoken~\cite{bonneau_science_2012, hasegawa_why_2021}, and only one considers urbanization differences within the same country~\cite{gilbert_network_2008}. 
Future work can continue to illuminate how security behavior changes based on sociodemographics other than national identity, such as within a country, in cultures that extend beyond nations, or WEIRD vs. non-WEIRD societies~\cite{linxenWERID2021}.

\paragraph{Race is an infrequently studied sociodemographic factor in research on security behaviors \opportunityRace.}
Race is a function of culture and was only studied in five papers.
Trends are difficult to ascertain because these papers investigated distinct behaviors and used different racial categorizations (we report using those papers' terminology).
With respect to security, prior work found that white people were more likely to take certain protective security actions than Asian Americans and Pacific Islanders as well as Black or African Americans, though American Indians and Alaska Natives were more likely than white people to use security settings~\cite{wash_too_2015}.
White people were more likely to solicit security advice from family and friends than Hispanic people~\cite{redmiles_secure_2016}.
With respect to privacy, while one paper found that racial minorities were more likely to publicly post on Snapchat~\cite{habib_snapchat_2019}, another found that compared to African Americans, white people were more likely to disclose distress on social media~\cite{zhang_distress_2021}.
One paper did not find racial differences in Facebook profile privacy settings~\cite{park_share_2018}. 

\subsection{Income}
\label{ss:results-income}

Income determines not only the financial resources that one has to spend, but it may also indirectly influence the time or energy that one can put towards security behaviors. 

\paragraph{More research is needed on relationships between income and security behaviors \opportunityIncome.}
Only three papers studied relationships between income and security behaviors: one found no differences in account sharing~\cite{park_share_2018}, while another found that people with lower incomes were more likely to adopt a combined set of security and privacy behaviors~\cite{zou_examining_2020}. 
People at different income levels learn from different sources; those with higher incomes were more likely to learn from school, work, or device prompts~\cite{redmiles_secure_2016}.


\section{Guidelines for Future Sociodemographic Research on Security Behaviors}
\label{sec:guidelines}
Our literature review documents a significant and growing interest in studying how sociodemographic factors relate to security behavior.
Based on our review, our own domain expertise, and sustained discussions amongst the research team, we developed guidelines to support strong and valuable research contributions. We iteratively refined these guidelines throughout our research process, including during our measurement study (see Section~\ref{sec:measurement}).
We offer these guidelines to assist researchers in both their research and reviewing process. However, we caution: the guidelines are not a checklist to guarantee quality work, there may be cases when they do not apply, and 
norms and best practices continually evolve.

\paragraph{Factor Selection.}
The selection of which sociodemographic factors to analyze should be done deliberately and stated as a research interest in the motivation (e.g., in the introductory section) for the work.
Many papers in our literature review
did not explicitly declare studying sociodemographic differences but presented correlations with sociodemographic factors in the results (see Section~\ref{ss:lit-results-focus}), which may indicate spurious correlations~\opportunityDeclare{}. 
Additionally, multiple studies are necessary to establish robust evidence of factor correlations, as demonstrated by the replication crisis in psychology research~\cite{maxwell_psychology_2015, shrout_psychology_2018}.

\begin{tcolorbox}[width=\linewidth, colback=white!95!black, boxrule=0.5pt, left=2pt,right=2pt,top=1pt,bottom=1pt]
\textbf{\guidelineFactorSelection{}: Identify at the beginning of the study the specific  sociodemographic factors, if any, you intend to study. } 
{If you investigate sociodemographic differences, commit to reporting the results even if they do not show differences, i.e., null results. Consider study pre-registration~\cite{cockburn_hark_2018}.}
\end{tcolorbox}

\paragraph{Group Selection.}
Within all sociodemographic factors, there are groups that are privileged or marginalized. 
We found many opportunities for research about different groups, e.g., groups marginalized by gender (see Section~\ref{ss:results-gender}) \opportunityGender~or race (see Section~\ref{ss:results-geography-race}) \opportunityRace.
Researchers choose to study a subset of groups for practical or other reasons.
If so, describe how the scope was chosen and how the sample studied relates to the broader population.

\begin{tcolorbox}[width=\linewidth, colback=white!95!black, boxrule=0.5pt, left=2pt,right=2pt,top=1pt,bottom=1pt]
\textbf{\guidelineGroupSelection{}:  
{Consider and justify which groups are included in or excluded from your study. }}
\end{tcolorbox}

\paragraph{Method Selection.}
Epistemic diversity allows researchers to explore a wider range of research questions.
Consider research methods that make different types of contributions~\cite{wobbrock_research_2016}, including but not limited to: quantitative, qualitative, or mixed methods~\cite{lazar_research_2017}; descriptive, experimental, or speculative; cross-sectional or longitudinal~\cite{gerken_longitudinal_2011, courage_longitudinal_2009}.
If relevant, consider causal inference methods~\cite{pearl_causal_2016, cunningham_2021_causal}.

Most papers in our literature review used statistical hypothesis testing, which is primarily valuable to identify factors for correlations but not causation.
Few papers in our literature review modeled sociodemographic factors as control factors (see Cho et al. as an exception~\cite{cho_networked_2016}).
Further, many papers we reviewed used self-report methods~\opportunitySelfReport, which are convenient for formative work but not suitable for establishing robust results.

\begin{tcolorbox}[width=\linewidth, colback=white!95!black, boxrule=0.5pt, left=2pt,right=2pt,top=1pt,bottom=1pt]
\textbf{\guidelineMethodSelection{}: 
{Consider using diverse research methods, acknowledging the benefits and limitations of each.}}
\end{tcolorbox}

\paragraph{Result Interpretation.}
When interpreting results, remember that complex factors could lead to any observed differences. 
Avoid ``essentializing'' (reducing individuals to assumed group characteristics) and over-generalizing findings. 
In interpreting results, state not only what can be implied from the results, but also what cannot: for example, ``We found a significant correlation between this factor and this behavior, which might be due to methodological choices or factors outside the scope of this study.'' 
This is particularly important for studies conducted on limited samples~\opportunityBalance{}.
\begin{tcolorbox}[width=\linewidth, colback=white!95!black, boxrule=0.5pt, left=2pt,right=2pt,top=1pt,bottom=1pt]
\textbf{G4: When sociodemographic differences are observed, exercise caution in describing the results.} Consider posing several causal interpretations for observed differences.
\end{tcolorbox}

\paragraph{Author Positionality.}
Weighing the advantages and disadvantages of disclosure~\cite{liang_embracing_2021}, if appropriate and safe to do so, include positionality statements in your work.
In some cases, the risks to researchers may not merit disclosure. Further, we caution against positionality statements that merely list identities without reflexivity as to how these identities influenced the research process.
When included thoughtfully, such statements provide context for readers about researcher motivations and the potential influence of researcher backgrounds.
For example, a majority of existing research in our literature review is U.S.-centric and studies people affiliated with universities~\opportunityEducation~and~\opportunityLocation, which is likely the result of the (undiscussed) positionality of researchers as primarily professors and graduate students at Western universities.

\begin{tcolorbox}[width=\linewidth, colback=white!95!black, boxrule=0.5pt, left=2pt,right=2pt,top=1pt,bottom=1pt]
\textbf{G5: Be aware of your own positionality and identity as a researcher and critically reflect on how it might affect your research questions, hypotheses, and interpretation of your findings~\cite{horton_critical_1999}.} 
\end{tcolorbox}


\section{Case Study: Measuring Sociodemographics and Security Behaviors on \platform}
\label{sec:measurement}

We now instantiate our guidelines in our own case study to concretely demonstrate their application for future 
researchers. We iteratively refined the guidelines in the process,
resulting in the version in Section~\ref{sec:guidelines}.

Unlike most prior work that uses self-reports, we leverage de-identified, aggregated log data to shed light on how users' real security behavior correlates with sociodemographic factors.
Security is often a secondary goal, so users may incorrectly recall actions and self-report based on social desirability~\cite{nederhof_methods_1985, krumpal_determinants_2013} or researcher demand~\cite{orne_social_1962}.
Thus, real-world security behavior offers high ecological validity and an important complement to self-report studies. 

\subsection{Measurement Methods}
\label{ss:measurement}
This study was conducted by combining de-identified, aggregated log data about security behavior with the results of a \numparticipants{}
respondent survey run on \platform{} in \numcountries{} countries during December 2019.
Respondents were recruited through both web and mobile interfaces via a message at the top of their social media feeds. The survey was translated into the respondent's local language by professional translators with native language proficiency.

Our \textit{dependent variables (DVs)} were four security behaviors:
    \textbf{Security settings visited} ($\pm$ 45 days of survey date), 
    \textbf{Security settings acted on} ($\pm$ 45 days of survey date, only among respondents who visited),
   \textbf{2FA enabled} (ever), and 
    \textbf{Stronger password} (i.e., those not yet identified as potentially more vulnerable to attack\footnote{See \url{https://www.facebook.com/help/124904560921566} for details on Facebook's password guidelines and \url{https://www.facebook.com/notes/760840091433907/} on identifying potentially vulnerable passwords.}).
    
Based on the available log data about \platform users, we chose six sociodemographic factors to study \guidelineBox{\guidelineFactorSelection{}}. These factors were: \textbf{Age}, \textbf{Gender} (binary\footnote{Due to cross-cultural differences in prevalence of non-binary gender reporting, we study only those who reported a binary gender to allow for interpretable comparisons across countries. As underscored in \opportunityGender, we encourage future work on those of non-binary genders.}), \textbf{Educational attainment}, \textbf{Geographic location} (16 countries in four regions), \textbf{Internet skill}, and \textbf{Technical knowledge}. Appendix~\ref{sec:appendix-factors} details these factors and how they were determined.
We also included four available \textit{independent variables} \textit{(IVs)} regarding \platform use based on the de-identified platform data:
    \textbf{Tenure} (how long the respondent had an account),
    \textbf{L30} (how many of the last 30 days the respondent had logged in to their account),
   \textbf{Time spent} (how much time the respondent spent on the platform over the last 30 days), and
    \textbf{Friend count} (number of social connections on the platform).

\paragraph{Analysis.}
We analyzed our data with logistic regression models because of the scale of our data \guidelineBox{\guidelineMethodSelection{}}. 
We weighted our sample to represent the population of the broader social media platform on age, gender, tenure, and L30 in order to maximize the generalizability of our results. 
To examine the relationship between security behavior and our independent variables, we constructed weighted logistic regression models, with security behavior as the boolean DV and the other variables listed above as the IVs (see Appendix~\ref{sec:appendix-regression}).
We also controlled for two interactions that had correlations with $\rho>0.3$: \textit{l$30*$time spent} --- there is a correlation between the number of days and amount of time spent on the platform --- and \textit{location$*$tenure} --- there is a correlation between geographic location and platform tenure since the platform was introduced to different markets at different times. Regression models were fit using 5-fold cross validation. The variance in AIC between the five folds was always less than 3\%.

\paragraph{Limitations.}
Our measurement study considers users of only one social media platform, although this platform is one of the largest and most diverse online platforms.
Though we studied users in \numcountries{} countries, this represents a minority of countries globally
\guidelineBox{\guidelineGroupSelection{}}.
Further, racial categories differ greatly by sociocultural context, which is why our measurement study across \numcountries{} countries did not study race \guidelineBox{\guidelineGroupSelection{}}.

\paragraph{Ethics.}
We analyze de-identified, aggregated log data of users on \platform{} who voluntarily completed survey data. There was no manipulation of any user’s experience, and no personal identifying information was used.
Our research procedures were vetted and approved through an internal review process.

\paragraph{Positionality.}

We echo our positionality statement from Section~\ref{ss:positionality} in conducting this measurement work \guidelineBox{\guidelineAuthorPositionality{}}. 
Additionally, we note that one author engaged in a paid collaboration with Meta, which allowed them to access and analyze the de-identified, aggregated log data.

\subsection{Measurement Results}
In interpreting our results, we emphasize that all findings describe only associations between sociodemographics and behaviors, and we do not make causal claims \guidelineBox{\guidelineResultInterpretation{}}.

\paragraph{Gender: On \platform, women were more likely than men to take action regarding security settings, but no gender differences were found with respect to password strength or use of 2FA.} We do not find significant differences in likelihood to \textit{visit} security settings, but women were more than 1.4 times as likely 
to \textit{action} security settings than men ($OR=1.44, p<.01$).
These results may support \trendGender~if actioning security settings is interpreted as an information protection behavior.
Given that other work on the same platform we study finds that people tend not to make a clear distinction between security and privacy~\cite{redmiles2019just}, women actioning security settings would align with other information protection behaviors.
We found no significant differences in likelihood to have a stronger password or use 2FA by gender.
While this null result could mean there is no relationship between gender and these behaviors, it could also mean that \platform{} users are unique in not having gender differences, but differences could be found in studies of users of other services.

\paragraph{Age: While older \platform{} users were less likely to visit their security settings, action their security settings, and use 2FA, those age 50+ were more likely to use stronger passwords.}  Compared to those aged 25-34, older adults were 
significantly less likely to \emph{view} their security settings, with the odds of those between 35-49 being 0.74 times as likely to visit their security settings ($OR_{35-49}=0.74, p_{35-49}<.05$) and those 50+ being 0.63 times as likely ($OR_{50+}=0.63, p_{50+}<.05$ ). 
We also found significant differences in their use security settings, with the odds of older adults \emph{actioning} their security settings being lower
 ($OR_{35-49}=0.55, p_{35-49}<.001$; $OR_{50+}=0.38, p_{50+}<.001$) and the odds of them using 2FA being lower, as well ($OR_{35-49}=0.79, p_{35-49}<.05$; $OR_{50+}=0.63, p_{50+}<.05$).
However, the odds of adults 50+ having a stronger password was higher ($OR_{50+}=2.08, p_{50+}<.05$).
These findings appear to support \trendAge, i.e., that older adults are more likely than younger to adopt security behaviors like passwords.

\paragraph{Education: On \platform{}, education levels were correlated with the likelihood of using 2FA.}
Compared to users with no post-secondary education, users with some college ($OR=7.14, p<.01$) or a bachelor's degree or more ($OR=5.40, p<.01$) were more likely to use 2FA.
However, education \emph{was not} correlated with visiting or actioning security settings or having a stronger password, in alignment with \trendEducation.

It is possible that users with higher educational levels had to previously comply with their institution's 2FA IT policy and thus were more likely to reengage with 2FA on \platform{}. 
Those with higher educations may also be more comfortable with computer systems and 
security tools like 2FA.
Future work can continue to investigate post-secondary institution's influence on 2FA adoption by comparing with users not affiliated with post-secondary institutions, towards \opportunityEducation.

\paragraph{Technical expertise: On \platform{}, technical expertise was correlated with stronger passwords and 2FA use.}
Technical knowledge of passwords was correlated with having a stronger password ($OR=1.88, p<.05$) and using 2FA ($OR=1.33, p<.05$), as was 
knowledge of the reaction feature on \platform{} ($OR=1.75, p<.05$; $OR=1.37, p<.01$for stronger password and 2FA, respectively).
Knowledge of QR codes was also correlated with greater use of 2FA ($OR=1.49, p<.001$), while knowledge of downloads was not correlated with any security behavior.
Since downloads are the oldest technology feature we asked about, the trends we find in our measurement study seem in alignment with our literature review, i.e., that technical expertise
correlates with increased likelihood to take secure actions \trendExpertise.

\paragraph{Internet skill: On \platform{}, internet skill was correlated with all behaviors except having a stronger password.}
Internet skill was correlated with visiting ($OR=1.41, p<.01$) and actioning ($OR=1.44, p<.05$) security settings as well as using 2FA ($OR=1.84, p<.001$).

\paragraph{Platform-specific use: Tenure on a platform was correlated with all security behaviors, while use in the past 30 days was not correlated with any.}
Platform tenure in years was correlated with all four security behaviors, specifically to be less likely to visit ($OR=0.95, p<.01$) or action ($OR=0.95, p<.05$) security settings, less likely to have a stronger password ($OR=0.91, p<.05$), but more likely to use 2FA ($OR=1.84, p<.001$). This may be due to those with longer standing accounts having already adjusted their settings and due to changes over time in password advice (those creating accounts earlier may have received less password education at the time of account creation). 
Friends and time spent were also positively correlated with use of 2FA ($OR=1.02, p<.01$; $OR=1.13, p<.05$), though use in the last 30 days was not correlated with any security behaviors.

\paragraph{Geography: On \platform{}, users in Africa, the Middle East, and Asian geographic markets differed significantly from the Western market in terms of security behavior.}
The odds of users in Asia visiting security settings were higher than users in the West ($OR=1.94, p<.05$), lower compared to the same group to have a stronger password ($OR=0.16, p<.001$), and no different for actioning security settings and using 2FA.
Users in Africa and the Middle East were less likely to have a stronger password ($OR=0.24, p<.05$), but other behaviors were not significantly different.
Users in Latin America were not significantly different from Western users on any of the four security behaviors we studied.
Geographic differences in our case study broadly align with \trendGeography{}, i.e., that geographic differences are significant but with unclear patterns \opportunityGeoMixed.


\section{Discussion}
\label{sec:discussion}

Having presented a systematization of knowledge of sociodemographics and security behaviors (Section~\ref{sec:lit-results}) and guidelines for researchers (Section~\ref{sec:guidelines}) and applied them in our own measurement study (Section~\ref{sec:measurement}), 
we now critically consider our \textit{lack} of knowledge in this space.

\subsection{The Missing ``Why?''}

This work reveals many correlations between sociodemographic factors and security behaviors, but little insight into \textit{why} these correlations exist.
The trends we synthesize and the opportunities we highlight begin to pose hypotheses for underlying causal relationships, but much work remains.
Without understanding why, interpreting results becomes arduous and different studies can yield seemingly contradictory results.
For example, when correlational studies find that one sociodemographic group adopts a security behavior less than another, is this the result of sample differences, different threat models, user interface designs that assumed one group as  ``default'' users~\cite{costanza_design_2020}, or some other reason?
Sociodemographic factors also do not exist in isolation but are correlated and influence each other; though this work did not investigate (reflecting papers in our literature review) intersectionality~\cite{crenshaw_intersectionality_2017}, only through understanding \textit{why} each identity influences behavior can intersectional analyses be conducted.

Drawing implications for interventions to change behavior when the \textit{why} is still missing is a tenuous proposition.
We can neither know what interventions might encourage adoption of security behaviors nor whether such interventions are necessary, desired, or even helpful. Worse, if we assume incorrectly, subsequent actions or discussions may have a negative impact, e.g., perpetuating gendered stereotypes about computer security and privacy behaviors~\cite{wei_stereotypes_23}. Recent related work on underlying causes of  differences in security threats, rather than on behaviors, takes initial steps toward a causality-focused framework, e.g., identifying higher-order factors like prominence and marginalization that put particular groups at higher risk of security threats~\cite{warford_sok_2022, sannon_privacy_2022}.

\subsection{Towards Answering ``Why?''}
\label{ss:disc-towards}
What should be next for this field of research on sociodemographics and computer security and privacy?
To close the knowledge gap, future work should explore not only \textit{what} differences exist among sociodemographic groups in security and privacy, but \textit{why} these differences exist.

\paragraph{Epistemic Diversity of Methods.}
Seeking to understand the causal relationships underlying sociodemographics and security behaviors cannot be achieved solely through quantitative methods.
That does not mean establishing correlations has no value; the field as a whole must grapple with \textit{why}, and individual papers provide incremental steps towards an answer.

In addition to the inferential methods used in the quantitative papers we analyzed, qualitative methods, e.g., in-depth interviews, observational studies, and ethnographies, can be used to explore the missing \textit{why}.
Such methods are increasingly used in security and privacy research to study the needs and practices of specific marginalized and vulnerable user groups but 
should also be used to draw out the underlying sociodemographic factors and their relationships to behavior. 
Especially by critically comparing privileged and marginalized groups, qualitative methods can assess existing hypotheses about causal relationships or pose new relationships and mechanisms of effect.

There are also other quantitative methods to consider beyond correlation and regression analyses. 
For example, structured equation modeling (SEM) involves constructing a model with causal relationships and statistically evaluating relationships as well as effect magnitudes~\cite{ullman_structural_2012}.
Other analyses include causal inference methods~\cite{pearl_causal_2009}, Bayesian methods~\cite{kruschke_doing_2014}, or quantitative meta-analyses.
Each method has strengths and limitations that future work can explore. 

\paragraph{Towards Social Theories.}
Overall, we recommend that security and privacy researchers learn from other fields that rely on social theories. 
Social theories, i.e., scientifically plausible principles that seek to explain certain phenomena by posing causal hypotheticals, pose richer explanations for how people behave, which
avoids essentializing a group of people.
For example, when women take fewer security measures than men, some might interpret this to mean that men are fundamentally better suited to security tasks.
Instead, women's choices may reflect systemic educational inequities, where women were discouraged from learning about technical topics, or other reasons.
Research must be careful to avoid attributing differences to innate group characteristics, e.g., racial essentializing~\cite{saini_superior_2019}.
Relevant social theories support robust interpretation when differences are found and can also indicate how a lack of difference can be meaningful.

Social theories from other fields can also help illuminate gaps in security behavior research, i.e., understudied factors that also merit study. 
Papers we analyzed focused most often on factors such as gender and age, but factors that have been more deeply studied in other fields and could inform security research include (dis)ability, marital status, religion, migration status, socioeconomic status, and race.

Finally, social theories facilitate critically challenging assumptions inherent in some perspectives on sociodemographics and security. This includes (1) questioning whether certain security behaviors are desirable for certain groups and in certain contexts since ``spending more time on security is not an inherent good''~\cite{herley2014more}. 
Further, (2) the security behaviors studied may not address (or be trusted to address) the needs of all communities, especially those most marginalized~\cite{mcdonald2021s, wilcox_infrastructuring_2023}, and (3) sociodemographic categorizations themselves, and the types of security behaviors studied, are not the only ways to organize the space and may not be the most salient to users.
As research continues to explore sociodemographic differences in security, incorporating theoretically informed inquiries presents the greatest opportunity to build on current methods and knowledge.


\section{Related Work}

\vspace{-0.1in}
\paragraph{Qualitative Studies of Marginalized Populations in Security and Privacy.}
A sizeable and growing body of literature investigates the experiences, behaviors, and needs of populations underrepresented in security and privacy research.
These works often overlap with sociodemographic factors, e.g., targets of intimate partner violence~\cite{tseng_care_2022} who are disproportionately women, refugees~\cite{simko_refugees_2018}, LGBTQ+ individuals~\cite{geeng_lesbians_2022}, and Muslim-American women~\cite{sambasivan_privacy_2018}.
Other studies investigate vulnerable populations due to their work, e.g., journalists~\cite{mcgregor_investigating_2015}, content creators~\cite{thomas_common_2022}, and sex workers~\cite{mcdonald2021s}.
These studies have been overwhelmingly qualitative, i.e., 
providing rich insights rather than quantitatively generalizable results.

\paragraph{Meta-Analyses of User Studies.}
The need for cross-study synthesis grows as the number of user studies of security behavior  increases, e.g., about
which methods are common~\cite{distler_systematic_2021}, or expert vs. non-expert users~\cite{kaur_human_2021}.
Prior meta-analyses also investigated marginalized~\cite{sannon_privacy_2022} and at-risk users~\cite{warford_sok_2022}, specifically developing unifying frameworks.
We focus on sociodemographics as the unifying frame because they are a powerful latent cause of differences; ultimately, marginalization relies on our contemporary, socially constructed sociodemographic categories.

Aside from a recent preprint investigating geographic diversity in security and privacy research~\cite{hasegawa_survey_2023}, we are unaware of other meta-reviews taking a sociodemographic lens, though sociodemographic meta-reviews in HCI are more common, e.g., culture~\cite{kamppuri_expanding_2006} as well as gender, race, and class~\cite{schlesinger_intersectional_2017}.


\section{Conclusion}
We broadly survey scholarship (\numpapers papers) 
that quantitatively studies sociodemographic factors and computer security behaviors, and we synthesize methods and results in a focused review of \numfocuspapers papers.
Taking a critical demography approach, we enumerate five trends in existing research and fifteen opportunities for future research (Table~\ref{tab:trends-opportunities}). 
We establish five guidelines for conducting quality sociodemographic research investigating security behaviors (Section~\ref{sec:guidelines}) and apply those guidelines in a case study of the real security behaviors of \numparticipants \platform users.
Taken together, this work documents the current state of knowledge on how people's identities relate to the security and privacy actions they take and charts new directions towards greater security, privacy, and equity.


\section{Acknowledgements}
We thank our reviewers and especially our shepherd for their helpful feedback.
We are also grateful to \alannah, \matthiasfassl, and the UW Security and Privacy Research Lab (including
\rachelhong, \alexandraemichael,~\christinayeung)
for insightful conversations on framing, methods, and impact.
We thank \maximilian~and \aleksei~for sharing an NDSS paper archive.
This work was supported in part by the U.S. National Science Foundation under Awards 2205171 and 2206950
and the Graduate Research Fellowship
Program (DGE-1746047). The fifth author did a portion of this work while working as a contractor for Meta.

\bibliographystyle{abbrv}
{\small
\bibliography{sociodemographics.bib}
}

\appendix

\section{Literature Review}
\subsection{Codebooks}
\label{sec:appendix-codebooks}
We developed three codebooks for our literature review and detail here the topics we coded for, labels within each codebook, as well as definitions and/or examples of each label.

\paragraph{Type of Security Behavior}: See Table~\ref{tab:counts-behaviors}.

\paragraph{Recruitment / Sample Characteristics}
\begin{itemize}
    \item \textbf{Representative}: successfully matched sample and population characteristics during recruitment
    \item \textbf{Balanced}: similar number of participants in factor groups for analysis, 
    including work that happened to have balanced samples
    \item \textbf{Limited}: sample characteristics were uncontrolled, e.g., snowball / convenience samples, data collected based on non-sociodemographic criteria and descriptively reports sociodemographic data
\end{itemize}

\paragraph{Study Methods}
\begin{itemize}
    \item \textbf{Self-report}: participants reporting security actions, e.g., interviews conducted in-person or via telephone, surveys
    \item \textbf{Measurement (Observational)}: scraped public data, log data (e.g., from a university, company), or data from software on participants' devices (e.g., mobile app or browser)
    \item \textbf{Measurement (Experimental)}: controlled condition or direct interaction involved, including non-lab study (e.g., MTurk experiment, experiment on social media) 
    or lab study (e.g., in-person, in-lab studies)
\end{itemize}

\subsection{Full List of Papers}
Tables~\ref{appendix:focus},~\ref{appendix:full1}, and~\ref{appendix:full2} show all papers in our literature review.

\section{Case Study}

\subsection{Sociodemographic Factors}
\label{sec:appendix-factors}

We provide here more details on the sociodemographic factors we study in our case study in Section~\ref{sec:measurement}. 
Based on the available log data about \platform users, we chose six sociodemographic factors to study:
\begin{itemize}
    \item \textbf{Age}, self-reported, bucketed into three groups; 25-34 (46.3\%), 35-49 (31.8\%), 50+ (21.9\%); min: 25, median: 36, mean: 40.14, max: 100
    \item \textbf{Gender}, gender (encoded as binary): women (43.5\%), men (56.5\%) 
    \item \textbf{Educational attainment}, self-reported, scaled per country (e.g., for Brazil, médio incompleto, superior completo, especialização), bucketed into three groups: high school equivalent or less (45.5\%), some college (23.6\%), BA or higher (30.8\%)
    \item \textbf{Geographic location}, one of \numcountries{} platform-inferred countries grouped into four regions: \textit{Western} (30.8\%): France, Italy, U.S., U.K.; \textit{Latin America} (17.4\%): Brazil, Mexico; \textit{Africa and Middle East} (22.4\%): Egypt, Kenya, Nigeria, Turkey; \textit{Asia} (29.4\%): India, Indonesia, Myanmar, Pakistan, Philippines, Vietnam
    \item \textbf{Internet skill}, measured via Web-Use Skills Index~\cite{hargittai2012succinct,hargittai2013new}, a standardized and validated self-report measure (min: 1.5, median: 3.75, mean: 3.66, max: 5).
    \item \textbf{Technical knowledge}, measured via Pew Research's password knowledge question~\cite{olmstead2017public} and three additional questions designed in the same style to assess familiarity with downloads, QR codes, and reacting to posts on the platform of study. Questions ask, ``Which of the following best describes'' and gives 5 answer options.
    (Password: 54.1\% correct\footnote{Pew Research~\cite{olmstead2017public} found that 75\% of U.S. survey respondents answered this question correctly; in our dataset 79.2\% of U.S. respondents did so.}; Download: 61.5\% correct; QR: 39.6\% correct; Reaction: 46.9\% correct)
\end{itemize}

\subsection{Regression Results}
\label{sec:appendix-regression}

Table~\ref{tab:reg-results} shows regression results for our case study (see Section~\ref{sec:measurement}).

\clearpage
\onecolumn
\begin{table}[tb]
\centering
\vspace{-0.4in}
\caption{All 47 papers in our focus dataset, i.e., from our seven selected conferences.}
\label{appendix:focus}
\scriptsize
\setlength{\tabcolsep}{2pt}
\begin{tabularx}{\textwidth}{rrp{.76\textwidth}r}
\toprule
\textbf{Year} & \textbf{Authors} & \textbf{Title}                                                               & \textbf{Conference}      \\ \midrule
\cellcolor{white}\textbf{2006}& Dhamija et al. &  Why Phishing Works & CHI\\
\rowcolor[HTML]{d9d9d9}
\cellcolor{white}\textbf{2007}& Sheng et al. &  Anti-Phishing Phil: The Design and Evaluation of a Game That Teaches People Not to Fall for Phish & SOUPS\\
\cellcolor{white}\textbf{2008}& Gilbert et al. &  The network in the garden: an empirical analysis of social media in rural life & CHI\\
\rowcolor[HTML]{d9d9d9}
\cellcolor{white}\textbf{2009}& Chiasson et al. &  Multiple password interference in text passwords and click-based graphical passwords & CCS\\
\cellcolor{white}& Kumaraguru et al. &  School of Phish: A Real-World Evaluation of Anti-Phishing Training & SOUPS\\
\rowcolor[HTML]{d9d9d9}
\cellcolor{white}\textbf{2010}& Shay et al. &  Encountering stronger password requirements: user attitudes and behaviors & SOUPS\\
\cellcolor{white}& Sheng et al. &  Who falls for phish?: a demographic analysis of phishing susceptibility and effectiveness of interventions & CHI\\
\rowcolor[HTML]{d9d9d9}
\cellcolor{white}& Stutzman et al. &  Friends only: examining a privacy-enhancing behavior in Facebook & CHI\\
\cellcolor{white}\textbf{2011}& Kaye & Self-reported password sharing strategies & CHI\\
\rowcolor[HTML]{d9d9d9}
\cellcolor{white}& Sotirakopoulos et al. &  On the challenges in usable security lab studies: lessons learned from replicating a study on SSL warnings & SOUPS\\
\cellcolor{white}& Wimberly et al. &  Using Fingerprint Authentication to Reduce System Security: An Empirical Study & IEEE S\&P\\
\rowcolor[HTML]{d9d9d9}
\cellcolor{white}\textbf{2012}& Bonneau et al. &  The science of guessing: analyzing an anonymized corpus of 70 million passwords & IEEE S\&P\\
\cellcolor{white}& Onarlioglu et al. &  Insights into User Behavior in Dealing with Internet Attacks & NDSS\\
\rowcolor[HTML]{d9d9d9}
\cellcolor{white}\textbf{2013}& Mazurek et al. &  Measuring Password Guessability for an Entire University & CCS\\
\cellcolor{white}\textbf{2014}& Chen et al. &  Exploring Internet Security Perceptions and Practices in Urban Ghana & SOUPS\\
\rowcolor[HTML]{d9d9d9}
\cellcolor{white}\textbf{2015}& Ion et al. &  ``...no one can hack my mind'': Comparing Expert and Non-Expert Security Practices & SOUPS\\
\cellcolor{white}& Jia et al. &  Risk-taking as a Learning Process for Shaping Teen's Online Information Privacy Behaviors & CSCW\\
\rowcolor[HTML]{d9d9d9}
\cellcolor{white}& Wash \& Rader & Too Much Knowledge? Security Beliefs and Protective Behaviors Among United States Internet Users & SOUPS\\
\cellcolor{white}\textbf{2016}& Cho et al. &  Networked Privacy Management in Facebook: A Mixed-Methods and Multinational Study & CSCW\\
\rowcolor[HTML]{d9d9d9}
\cellcolor{white}& Harbach et al. &  Keep on Lockin' in the Free World: A Multi-National Comparison of Smartphone Locking & CHI\\
\cellcolor{white}& Redmiles et al. &  How I Learned to be Secure: a Census-Representative Survey of Security Advice Sources and Behavior & CCS\\
\rowcolor[HTML]{d9d9d9}
\cellcolor{white}\textbf{2017}& Bonné et al. &  Exploring decision making with Android's runtime permission dialogs using in-context surveys & SOUPS\\
\cellcolor{white}& Fiesler et al. &  What (or Who) Is Public?: Privacy Settings and Social Media Content Sharing & CSCW\\
\rowcolor[HTML]{d9d9d9}
\cellcolor{white}& Hoyle et al. &  Viewing the Viewers: Publishers' Desires and Viewers' Privacy Concerns in Social Networks & CSCW\\
\cellcolor{white}& Pearman et al. &  Let's Go in for a Closer Look: Observing Passwords in Their Natural Habitat & CCS\\
\rowcolor[HTML]{d9d9d9}
\cellcolor{white}\textbf{2018}& Das et al. &  Breaking! A Typology of Security and Privacy News and How It's Shared & CHI\\
\cellcolor{white}& Machuletz & Webcam Covering as Planned Behavior & CHI\\
\rowcolor[HTML]{d9d9d9}
\cellcolor{white}& Redmiles et al. &  Examining the Demand for Spam: Who Clicks? & CHI\\
\cellcolor{white}& Sharif et al. &  Predicting Impending Exposure to Malicious Content from User Behavior & CCS\\
\rowcolor[HTML]{d9d9d9}
\cellcolor{white}\textbf{2019}& Habib et al. &  Impact of Contextual Factors on Snapchat Public Sharing & CHI\\
\cellcolor{white}\textbf{2020}& Coopamootoo & Usage Patterns of Privacy-Enhancing Technologies & CCS\\
\rowcolor[HTML]{d9d9d9}
\cellcolor{white}& Zou et al. &  Examining the Adoption and Abandonment of Security, Privacy, and Identity Theft Protection Practices & CHI\\
\cellcolor{white}\textbf{2021}& Hasan et al. &  Your Photo is so Funny that I don't Mind Violating Your Privacy by Sharing it 
& CHI\\
\rowcolor[HTML]{d9d9d9}
\cellcolor{white}& Zhang et al. &  Distress Disclosure across Social Media Platforms during the COVID-19 Pandemic
& CHI\\
\cellcolor{white}\textbf{2023}& Bouma-Sims et al. &  A US-UK Usability Evaluation of Consent Management Platform Cookie Consent Interface Design on Desktop and Mobile & CHI\\
\rowcolor[HTML]{d9d9d9}
\cellcolor{white}& Munyendo et al. &  I just stopped using one and started using the other: Motivations Techniques and Challenges When Switching Password Managers & CCS\\

\bottomrule

\end{tabularx}
\end{table}

\vspace{-0.5in}
\begin{table}[t!]
\centering
\caption{The remaining papers in our full dataset (1999-2014).}
\label{appendix:full1}
\scriptsize
\setlength{\tabcolsep}{2pt}
\begin{tabularx}{\textwidth}{rrp{.64\textwidth}p{.2\textwidth}}
\toprule
\textbf{Year} & \textbf{Authors} & \textbf{Title}  & \textbf{Venue}      \\ \midrule
\rowcolor[HTML]{d9d9d9}
\cellcolor{white}\textbf{2004} & Milne \& Culnan  &  Strategies for reducing online privacy risks: Why consumers read (or don'tread) online privacy notices &Jrnl. of Interactive Marketing \\
\cellcolor{white}     & Milne et al.  &  Consumers' Protection of Online Privacy and Identity. &Jrnl. of Consumer Affairs \\
\rowcolor[HTML]{d9d9d9}
\cellcolor{white}\textbf{2005} & Youn  &  Teenagers' Perceptions of Online Privacy and Coping Behaviors: A Risk–Benefit Appraisal Approach & Jrnl. of Broadcasting \& E. Media \\
\cellcolor{white}\textbf{2007} & Grimes et al.  &  Email end users and spam: relations of gender and age group to attitudes and actions Computers in & Human Behavior \\
\rowcolor[HTML]{d9d9d9}
\cellcolor{white}     & Jagatic et al.  &  Social phishing & CACM \\
\cellcolor{white}     & Kumaraguru et al.  &  Getting Users to Pay Attention to Anti-Phishing Education: Evaluation of Retention and Transfer & APWG \\
\rowcolor[HTML]{d9d9d9}
\cellcolor{white}     & Kuo et al.  &  Assessing Gender Differences in Computer Professionals' Self-Regulatory Efficacy Concerning Info. Privacy Practices &Jrnl. of Business Ethics \\
\cellcolor{white}\textbf{2008} & Bailey et al.  &  Analysis of Student Vulnerabilities to Phishing. & AMCIS \\
\rowcolor[HTML]{d9d9d9}
\cellcolor{white}     & Hazari et al.  &  An Empirical Investigation of Factors Influencing Information Security Behavior &Jrnl. of Info. Privacy \& Security \\
\cellcolor{white}     & Lewis et al.  &  The Taste for Privacy: An Analysis of College Student Privacy Settings inan Online Social Network &Jrnl. of Comp. Mediated Comm. \\
\rowcolor[HTML]{d9d9d9}
\cellcolor{white}     & Youn \& Hall  &  Gender and Online Privacy among Teens: Risk Perception, Privacy Concerns, and Protection Behaviors & Cyber Psychology \& Behavior \\
\cellcolor{white}\textbf{2009} & Dinev et al.  &  User behaviour towards protective information technologies: the role ofnational cultural differences & Information Systems Journal \\
\rowcolor[HTML]{d9d9d9}
\cellcolor{white}     & Fogel \& Nehmad  &  Internet social network communities: Risk taking, trust, and privacy concerns&  Computers in Human Behavior \\
\cellcolor{white}     & Milne et al.  &  Toward an Understanding of the Online Consumer's Risky Behavior and Protection Practices &Jrnl. of Consumer Affairs \\
\rowcolor[HTML]{d9d9d9}
\cellcolor{white}\textbf{2010} & Brandtzæg  &  Too Many Facebook “Friends”? Content Sharing and Sociability Versus the Need for Privacy in Social Network Sites & Jrnl. of HCI \\
\cellcolor{white}     & Durand  &  A Comparative Study of Self-Disclosure in Face-to-Face and Email Communication Between Americans and China & N/A (thesis) \\
\rowcolor[HTML]{d9d9d9}
\cellcolor{white}     & Hoy \& Milne  &  Gender Differences in Privacy-Related Measures for Young Adult Facebook Users & Jrnl. of Interactive Advertising \\
\cellcolor{white}     & Posey et al.  &  Proposing the online community self-disclosure model 
& Euro. Jrnl. of Information Systems \\
\rowcolor[HTML]{d9d9d9}
\cellcolor{white}     & Siripukdee et al.  &  Empirical Analysis of Human-related Problems on Information Security in Cross-cultural Environment 
& Japan Society for Info. \& and Mgmt. \\
\cellcolor{white}     & Wright et al.  &  Where Did They Go Right? Understanding the Deception in Phishing Communications & Group Decision and Negotiation \\
\rowcolor[HTML]{d9d9d9}
\cellcolor{white}\textbf{2011} & Kruger et al.  &  An assessment of the role of cultural factors in information security awareness & Information SecuritySouth Africa \\
\cellcolor{white}     & Lomo-David et al.  &  University Students Computer Security Practices in Two Developing Nations: A Comparative Analysis & SHSU General Business Conference \\
\rowcolor[HTML]{d9d9d9}
\cellcolor{white}     & Lowry et al.  &  Privacy Concerns Versus Desire for Interpersonal Awareness in Driving the Use of Self-Disclosure Technologies 
&Jrnl. of Management Info. Systems \\
\cellcolor{white}     & Maier et al.  &  An Assessment of Overt Malicious Activity Manifest in Residential Networks & DIMVA \\
\rowcolor[HTML]{d9d9d9}
\cellcolor{white}     & Special et al.  &  Self-disclosure and student satisfaction with Facebook Computers in & Human Behavior \\
\cellcolor{white}\textbf{2012} & Krasnova et al.  &  Self-disclosure and Privacy Calculus on Social Networking Sites: The Roleof Culture & BISE \\
\rowcolor[HTML]{d9d9d9}
\cellcolor{white}     & Madden  &  Privacy management on social media sites & Pew \\
\cellcolor{white}     & Mohebzada et al.  &  Phishing in a university community: Two large scale phishing experiments & IIT \\
\rowcolor[HTML]{d9d9d9}
\cellcolor{white}     & Tufekci  &  Youth and Privacy in Networked Publics: Active and Complex Engagement & ICWSM \\
\cellcolor{white}\textbf{2013} & Halevi et al.  &  A pilot study of cyber security and privacy related behavior and personality traits & WWW \\
\rowcolor[HTML]{d9d9d9}
\cellcolor{white}     & Litt  &  Understanding social network site users' privacy tool use Computers in & Human Behavior \\
\cellcolor{white}     & Madden et al.  &  Teens, Social Media, and Privacy & Pew \\
\rowcolor[HTML]{d9d9d9}
\cellcolor{white}     & Park  &  Digital Literacy and Privacy Behavior Online & Communication Research \\
\cellcolor{white}     & Rainie et al.  &  Anonymity, Privacy, and Security Online & Pew \\
\rowcolor[HTML]{d9d9d9}
\cellcolor{white}\textbf{2014} & Alseadoon  &  The Impact of Users' Characteristics on Their Ability to Detect Phishing Emails & N/A (thesis) \\
\cellcolor{white}     & Baek et al.  &  My privacy is okay, but theirs is endangered: Why comparative optimism matters in online privacy concerns & Computers in Human Behavior \\
\rowcolor[HTML]{d9d9d9}
\cellcolor{white}     & Blank et al.  &  A New Privacy Paradox: Young People and Privacy on Social Network Sites  & ASA \\
\bottomrule

\end{tabularx}
\end{table}
\newpage

\begin{table}[t!]
\centering
\vspace{-0.3in}
\caption{The remaining papers in our full dataset (2014-2023).}
\label{appendix:full2}
\scriptsize
\setlength{\tabcolsep}{2pt}
\begin{tabularx}{\textwidth}{rrp{.67\textwidth}p{.17\textwidth}}
\toprule
\textbf{Year} & \textbf{Authors} & \textbf{Title}                                                               & \textbf{Venue}      \\ \midrule
\cellcolor{white}\textbf{2014} & Tembe et al.  &  Phishing in international waters 
& HotSoS \\
\rowcolor[HTML]{d9d9d9}
\cellcolor{white}     & Vanderhoven et al.  &  How Safe Do Teenagers Behave on Facebook? An Observational Study & PLoS One \\
\cellcolor{white}\textbf{2015} & Anderson et al.  &  Neural correlates of gender differences and color in distinguishing security warnings and legitimate websites 
&Jrnl. of Cybersecurity \\
\rowcolor[HTML]{d9d9d9}
\cellcolor{white}     & Halevi et al.  &  Spear-Phishing in the Wild 
& SSRN \\
\cellcolor{white}     & Marshall et al.  &  Social networking websites in India and the United States: A cross-national comparison of online privacy and communication & Issues in IS \\
\rowcolor[HTML]{d9d9d9}
\cellcolor{white}     & Park  &  Do men and women differ in privacy? Gendered privacy and (in)equality in the Internet & Computers in Human Behavior \\
\cellcolor{white}     & Pattinson et al.  &  Factors that Influence Information Security Behavior: An Australian Web Based Study & HAS \\
\rowcolor[HTML]{d9d9d9}
\cellcolor{white}     & Posey et al.  &  The Impact of Organizational Commitment on Insiders' Motivation to Protect Organizational Information Assets & Jrnl. of Management Information Systems \\
\cellcolor{white}     & Whitty et al.  &  Individual Differences in Cyber Security Behaviors: An Examination of Who Is Sharing Passwords & Cyber Psychology, Behavior, and Social Networking \\
\rowcolor[HTML]{d9d9d9}
\cellcolor{white}\textbf{2016} & Aviv et al.  &  Analyzing the Impact of Collection Methods and Demographics for Android's Pattern Unlock & USEC \\
\cellcolor{white}     & Bertenthal  &  Attention and Past Behavior, not Security Knowledge, Modulate Users' Decisions to Login to Insecure Websites & ICS \\
\rowcolor[HTML]{d9d9d9}
\cellcolor{white}     & Chen \& Zahedi  &  Individuals' Internet Security Perceptions and Behaviors: Polycontextual Contrasts Between the United States and China & MIS Quarterly \\
\cellcolor{white}     & Halevi et al.  &  Cultural and psychological factors in cyber-security II & WAS \\
\rowcolor[HTML]{d9d9d9}
\cellcolor{white}     & Iuga et al.  &  Baiting the hook: factors impacting susceptibility to phishing attacks & Human-centric Computing and Info. Sciences \\
\cellcolor{white}     & Kezer et al.  &  Age differences in privacy attitudes, literacy and privacy management on Facebook &Jrnl. of Psychosocial Research on Cyberspace \\
\rowcolor[HTML]{d9d9d9}
\cellcolor{white}     & Malik et al.  &  Privacy and trust in Facebook photo sharing: Age and gender differences & Program \\
\cellcolor{white}     & Petrie et al.  &  Cultural and Gender Differences in Password Behaviors: Evidence from China, Turkey and the UK & NordiCHI \\
\rowcolor[HTML]{d9d9d9}
\cellcolor{white}     & Reed et al.  &  Thumbs up for privacy?: Differences in online self-disclosure behavior across national cultures & Social Science Research \\
\cellcolor{white}     & Sonnenschein et al.  &  Gender Differences in Mobile Users' IT Security Appraisals and Protective Actions: Findings from a Mixed-Method Study & ISIC \\
\rowcolor[HTML]{d9d9d9}
\cellcolor{white}     & Tsay-Vogel et al.  &  Social media cultivating perceptions of privacy 
& New Media \& Society \\
\cellcolor{white}\textbf{2017} & Anwar et al.  &  The impact of collectivism and psychological ownership on protection motivation: A cross-cultural examination & Computers in Human Behavior \\
\rowcolor[HTML]{d9d9d9}
\cellcolor{white}     & Büchi et al.  &  Caring is not enough: the importance of Internet skills for online privacy protection & ICS \\
\cellcolor{white}     & Butavicius et al.  &  Understanding susceptibility to phishing emails: Assessing the impact of individual differences and culture & HAISA \\
\rowcolor[HTML]{d9d9d9}
\cellcolor{white}     & Gavett et al.  &  Phishing suspiciousness in older and younger adults: The role of executive functioning & PLoS One \\
\cellcolor{white}     & Ifinedo et al.  &  Effects of Organization Insiders' Self-Control and Relevant Knowledge on Participation in Information Systems Security Deviant Behavior & SIGMIS-CPR \\
\rowcolor[HTML]{d9d9d9}
\cellcolor{white}     & Sarno et al.  &  Who are Phishers luring?: A Demographic Analysis of Those Susceptible to Fake Emails & Human Factors and Ergonomics Society \\
\cellcolor{white}\textbf{2018} & Alohali et al.  &  Identifying and predicting the factors affecting end-users' risk-taking behavior &Jrnl. of Info. \& Comp. Security \\
\rowcolor[HTML]{d9d9d9}
\cellcolor{white}     & Cain et al.  &  An exploratory study of cyber hygiene behaviors and knowledge & Jrnl. of Information Security and Applications \\
\cellcolor{white}     & Diaz et al.  &  Phishing in an Academic Community: A Study of User Susceptibility and Behavior & ArXiV \\
\rowcolor[HTML]{d9d9d9}
\cellcolor{white}     & Farinosi \& Taipale  &  Who Can See My Stuff? Online Self-Disclosure and Gender Differences on Facebook & OBS \\
\cellcolor{white}     & Griffin  &  A Demographic Analysis to Determine User Vulnerability among Several Categories of Phishing Attacks & N/A (thesis) \\
\rowcolor[HTML]{d9d9d9}
\cellcolor{white}     & Lévesque et al.  &  Technological and Human Factors of Malware Attacks: A Computer Security Clinical Trial Approach & TOPS \\
\cellcolor{white}     & McGill et al.  &  Gender Differences in Information Security Perceptions and Behaviour & ACIS \\
\rowcolor[HTML]{d9d9d9}
\cellcolor{white}     & Menard et al.  &  The impact of collectivism and psychological ownership on protection motivation: A cross-cultural examination & Computers \& Security \\
\cellcolor{white}     & Millham et al.  &  Managing the virtual boundaries: Online social networks, disclosure, and privacy behaviors & New Media \& Society \\
\rowcolor[HTML]{d9d9d9}
\cellcolor{white}     & Redmiles  &  Net Benefits: Digital Inequities in Social Capital, Privacy Preservation, and Digital Parenting Practices of U.S. Social Media Users & ICWSM \\
\cellcolor{white}\textbf{2019} & Dev et al.  &  Personalized WhatsApp Privacy: Demographic and Cultural Influences on Indian and Saudi Users & SSRN \\
\rowcolor[HTML]{d9d9d9}
\cellcolor{white}     & Lin et al.  &  Susceptibility to Spear-Phishing Emails: Effects of Internet User Demographics and Email Content & ToCHI \\
\cellcolor{white}     & Ndibwile et al.  &  A Demographic Perspective of Smartphone Security and Its Redesigned Notifications &Jrnl. of Information Processing \\
\rowcolor[HTML]{d9d9d9}
\cellcolor{white}     & Shappie et al.  &  Personality as a Predictor of Cybersecurity Behavior & Psychology of Popular Media Culture \\
\cellcolor{white}\textbf{2019} & Tifferet et al.  &  Gender differences in privacy tendencies on social network sites: A meta-analysis & Computers in Human Behavior \\
\rowcolor[HTML]{d9d9d9}
\cellcolor{white}     & Breitinger et al.  &  A survey on smartphone user's security choices, awareness and education & Computers \& Security \\
\cellcolor{white}     & Epstein et al.  &  Markers of Online Privacy Marginalization: Empirical Examination of Socioeconomic Disparities in Social Media Privacy Attitudes, Literacy, and Behavior & Social Media + Society \\
\rowcolor[HTML]{d9d9d9}
\cellcolor{white}     & Herbert et al.  &  Differences in IT Security Behavior and Knowledge of Private Users in Germany & Wirtschaftsinformatik \\
\cellcolor{white}     & Li et al.  &  Experimental Investigation of Demographic Factors Related to Phishing Susceptibility & HICCS \\
\rowcolor[HTML]{d9d9d9}
\cellcolor{white}     & Liu et al.  &  Effects of Demographic Factors on Phishing Victimization in the Workplace & PACIS \\
\cellcolor{white}     & Oghazi et al.  &  User self-disclosure on social network sites: A cross-cultural study on Facebook's privacy concepts & Jrnl. of Business Research \\
\rowcolor[HTML]{d9d9d9}
\cellcolor{white}     & Sombatruang et al.  &  Attributes affecting user decision to adopt a Virtual Private Network (VPN) app & ICICS \\
\cellcolor{white}     & Thao et al.  &  Human Factors in Homograph Attack Recognition & ANCS \\
\rowcolor[HTML]{d9d9d9}
\cellcolor{white}     & Zwilling et al.  &  Cyber Security Awareness, Knowledge and Behavior: A Comparative Study &Jrnl. of Comp. Info. Systems \\
\cellcolor{white}\textbf{2021} & Abroshan et al.  &  COVID-19 and Phishing: Effects of Human Emotions, Behavior, and Demographics on the Success of Phishing Attempts During the Pandemic & IEEE Access \\
\rowcolor[HTML]{d9d9d9}
\cellcolor{white}     & Abroshan et al.  &  Phishing Happens Beyond Technology 
& IEEE Access \\
\cellcolor{white}     & Bhagavatula et al.  &  What breach? Measuring online awareness of security incidents by studying real-world browsing behavior & EuroUSEC \\
\rowcolor[HTML]{d9d9d9}
\cellcolor{white}     & Boerman et al.  &  Exploring Motivations for Online Privacy Protection Behavior: Insights From Panel & Data Comm. Research \\
\cellcolor{white}     & Greitzer et al.  &  Experimental Investigation of Technical and Human Factors Related to Phishing Susceptibility & ACM Transactions on Social Computing \\
\rowcolor[HTML]{d9d9d9}
\cellcolor{white}     & Grobler et al.  &  The importance of social identity on password formulations &  Personal and Ubi. Comp. \\
\cellcolor{white}     & Kennison et al.  &  Who creates strong passwords when nudging fails & Computers in Human Behavior \\
\rowcolor[HTML]{d9d9d9}
\cellcolor{white}     & Mai et al.  &  Cyber Security Awareness and Behavior of Youth in Smartphone Usage: A Comparative Study between University Students in Hungary and Vietnam & Acta Polytechnica Hungarica \\
\cellcolor{white}     & Morrison  &  Understanding U.S. Employees' Personality Traits for Phishing Emails Prevention: A Quantitative Study & N/A (thesis) \\
\rowcolor[HTML]{d9d9d9}
\cellcolor{white}     & Ouytsel  &  The prevalence and motivations for password sharing practices and intrusive behaviors among early adolescents' best friendships – A mixed-methods study & Telematics and Informatics \\
\cellcolor{white}     & Roberts  &  Does Digital Native Status Impact End-User Antivirus Usage? & Jrnl. of Comp. Net. \& Comm. \\
\rowcolor[HTML]{d9d9d9}
\cellcolor{white}\textbf{2022} & Frank et al.  &  Contextual drivers of employees' phishing susceptibility: Insights from a field study & Decision Support Systems \\
\cellcolor{white}\textbf{2023} & Du et al.  &  Phishing: Gender Differences in Email Security Perceptions and Behaviors & Info. Sys. and Comput. Academic Professionals \\
\bottomrule
\end{tabularx}
\end{table}

\begin{table*}
\caption{\small{Regression results for the relationships between security behaviors (first row, output variables) and sociodemographic factors and platform metrics (first column, input factors). Each column represents the output of one regression model. Numeric cells list the odds ratio (OR) and the 95\% confidence interval. Significance of OR: $p<0.05$ = \colorbox{gray!30}{*}, $p<0.01$ = \colorbox{gray!30}{**}, and $p<0.001$ = \colorbox{gray!30}{***}.} LATAM = Latin America, AME = Africa and Middle East, Edu. = Education, SC = some college, BA+ = Bachelor's degree or more, Tech. = Technical, Know. = Knowledge, L30 = Use (Past 30 days)}
\label{tab:reg-results}
\centering
\small
\renewcommand{\arraystretch}{1.7}
\newcolumntype{Y}{>{\centering\arraybackslash}X}
\begin{tabularx}{\textwidth}{p{4cm} Y Y Y Y}
\toprule
    & \textbf{Visit \newline Security Settings} 
    & \textbf{Action \newline Security Settings}  
    & \textbf{Stronger \newline Password} 
    & \textbf{Use 2FA}  \\
\midrule
(Intercept) 
    & \cellcolor{gray!30}\shortstack{0.02*** \\ $[0, 0.05]$ } 
    & \cellcolor{gray!30}\shortstack{0.02*** \\ $[0, 0.06]$ } 
    & \cellcolor{gray!30}\shortstack{192.25*** \\ $[29.13, 1268.55]$ } 
    & \cellcolor{gray!30}\shortstack{0*** \\ $[0, 0]$ } \\
Age (35-49) 
    & \cellcolor{gray!30}\shortstack{0.74* \\ $[0.59, 0.94]$ } 
    & \cellcolor{gray!30}\shortstack{0.55*** \\ $[0.44, 0.70]$ } 
    & \shortstack{1.18 \\ $[0.64, 2.17]$ } 
    & \cellcolor{gray!30}\shortstack{0.79* \\ $[0.64, 0.96]$ } \\
Age (50+) 
    & \cellcolor{gray!30}\shortstack{0.63* \\ $[0.42, 0.95]$ } 
    & \cellcolor{gray!30}\shortstack{0.38*** \\ $[0.23, 0.63]$ } 
    & \cellcolor{gray!30}\shortstack{2.08* \\ $[1.07, 4.03]$ } 
    & \cellcolor{gray!30}\shortstack{0.63* \\ $[0.43, 0.92]$ } \\
Gender (woman) 
    & \shortstack{1.20 \\ $[0.97, 1.49]$ } 
    & \cellcolor{gray!30}\shortstack{1.44** \\ $[1.14, 1.82]$ } 
    & \shortstack{1.55 \\ $[0.86, 2.80]$ } 
    & \shortstack{0.88 \\ $[0.73, 1.06]$ } \\
Location (LATAM) 
    & \shortstack{0.90 \\ $[0.56, 1.44]$ } 
    & \shortstack{0.90 \\ $[0.53, 1.55]$ } 
    & \shortstack{0.64 \\ $[0.20, 2.05]$ } 
    & \shortstack{0.85 \\ $[0.42, 1.71]$ } \\
\midrule
Location (AME) 
    & \shortstack{0.87 \\ $[0.52, 1.46]$ } 
    & \shortstack{0.73 \\ $[0.40, 1.35]$ } 
    & \cellcolor{gray!30}\shortstack{0.24* \\ $[0.08, 0.71]$ } 
    & \shortstack{1.06 \\ $[0.55, 2.03]$ } \\
Location (Asia) 
    & \cellcolor{gray!30}\shortstack{1.94* \\ $[1.15, 3.28]$ } 
    & \shortstack{1.61 \\ $[0.87, 2.99]$ } 
    & \cellcolor{gray!30}\shortstack{0.16*** \\ $[0.06, 0.45]$ } 
    & \shortstack{1.44 \\ $[0.68, 3.02]$ } \\
Edu. (SC)  
    & \shortstack{1.25 \\ $[0.41, 3.79]$ } 
    & \shortstack{1.25 \\ $[0.32, 4.77]$ } 
    & \shortstack{0.57 \\ $[0.04, 8.41]$ } 
    & \cellcolor{gray!30}\shortstack{7.14** \\ $[2.14, 23.85]$ } \\
Edu. (BA+) 
    & \shortstack{1.36 \\ $[0.42, 4.39]$ } 
    & \shortstack{1.39 \\ $[0.37, 5.16]$ } 
    & \shortstack{0.89 \\ $[0.15, 5.30]$ } 
    & \cellcolor{gray!30}\shortstack{5.40** \\ $[1.74, 16.72]$ } \\
Internet Skill 
    & \cellcolor{gray!30}\shortstack{1.41** \\ $[1.12, 1.78]$ } 
    & \cellcolor{gray!30}\shortstack{1.44* \\ $[1.09, 1.90]$ } 
    & \shortstack{1.10 \\ $[0.79, 1.53]$ } 
    & \cellcolor{gray!30}\shortstack{1.84*** \\ $[1.42, 2.39]$ } \\
\midrule
Tech. Know. (Download) 
    & \shortstack{1.02 \\ $[0.81, 1.29]$ } 
    & \shortstack{0.87 \\ $[0.68, 1.11]$ } 
    & \shortstack{0.90 \\ $[0.49, 1.65]$ } 
    & \shortstack{0.83 \\ $[0.66, 1.04]$ } \\
Tech. Know. (Password) 
    & \shortstack{0.97 \\ $[0.77, 1.23]$ } 
    & \shortstack{1.15 \\ $[0.89, 1.49]$ } 
    & \cellcolor{gray!30}\shortstack{1.88* \\ $[1.09, 3.23]$ } 
    & \cellcolor{gray!30}\shortstack{1.33* \\ $[1.01, 1.74]$ } \\
Tech. Know. (QR) 
    & \shortstack{1.14 \\ $[0.92, 1.42]$ } 
    & \shortstack{1.15 \\ $[0.90, 1.47]$ } 
    & \shortstack{1.64 \\ $[0.80, 3.36]$ } 
    & \cellcolor{gray!30}\shortstack{1.49*** \\ $[1.19, 1.87]$ } \\
Tech. Know. (Reaction) 
    & \shortstack{1.12 \\ $[0.90, 1.39]$ } 
    & \shortstack{1.24 \\ $[0.97, 1.58]$ } 
    & \cellcolor{gray!30}\shortstack{1.75* \\ $[1.02, 3.00]$ } 
    & \cellcolor{gray!30}\shortstack{1.37** \\ $[1.11, 1.70]$ } \\
Platform Tenure (Years) 
    & \cellcolor{gray!30}\shortstack{0.95** \\ $[0.91, 0.99]$ } 
    & \cellcolor{gray!30}\shortstack{0.95* \\ $[0.91, 0.99]$ } 
    & \cellcolor{gray!30}\shortstack{0.91* \\ $[0.84, 1.00]$ } 
    & \cellcolor{gray!30}\shortstack{1.08*** \\ $[1.04, 1.13]$ } \\
\midrule
Friends 
    & \shortstack{1.00 \\ $[0.99, 1.01]$ } 
    & \shortstack{1.00 \\ $[0.99, 1.01]$ } 
    & \shortstack{1.01 \\ $[0.98, 1.04]$ } 
    & \cellcolor{gray!30}\shortstack{1.02** \\ $[1.00, 1.03]$ } \\
Use (Past 30 days) 
    & \shortstack{1.00 \\ $[0.98, 1.01]$ } 
    & \shortstack{0.99 \\ $[0.97, 1.00]$ } 
    & \shortstack{0.97 \\ $[0.93, 1.01]$ } 
    & \shortstack{1.01 \\ $[0.99, 1.03]$ } \\
Time Spent 
    & \shortstack{1.01 \\ $[1.00, 1.18]$ } 
    & \shortstack{1.03 \\ $[0.94, 1.12]$ } 
    & \shortstack{0.90 \\ $[0.63, 1.29]$ } 
    & \cellcolor{gray!30}\shortstack{1.13* \\ $[1.02, 1.24]$ } \\
Edu. (SC) * Internet Skill 
    & \shortstack{0.91 \\ $[0.68, 1.22]$ } 
    & \shortstack{0.89 \\ $[0.63, 1.26]$ } 
    & \shortstack{1.30 \\ $[0.65, 2.60]$ } 
    & \cellcolor{gray!30}\shortstack{0.63** \\ $[0.47, 0.85]$ } \\
Edu. (BA+) * Internet Skill 
    & \shortstack{0.88 \\ $[0.65, 1.18]$ } 
    & \shortstack{0.85 \\ $[0.60, 1.19]$ } 
    & \shortstack{0.92 \\ $[0.56, 1.50]$ } 
    & \cellcolor{gray!30}\shortstack{0.71* \\ $[0.53, 0.95]$ } \\
\midrule
L30 * Time Spent 
    & \shortstack{1.00 \\ $[1.00, 1.00]$ } 
    & \shortstack{1.00 \\ $[1.00, 1.01]$ } 
    & \shortstack{1.01 \\ $[0.99, 1.02]$ } 
    & \shortstack{1.00 \\ $[0.99, 1.00]$ } \\
LATAM * Platform Tenure 
    & \shortstack{1.03 \\ $[0.97, 1.10]$ } 
    & \shortstack{1.05 \\ $[0.98, 1.12]$ } 
    & \shortstack{0.99 \\ $[0.87, 1.14]$ } 
    & \shortstack{0.95 \\ $[0.87, 1.03]$ } \\
AME * Platform Tenure 
    & \shortstack{1.05 \\ $[0.99, 1.11]$ } 
    & \cellcolor{gray!30}\shortstack{1.09* \\ $[1.02, 1.17]$ } 
    & \shortstack{1.08 \\ $[0.96, 1.22]$ } 
    & \shortstack{1.00 \\ $[0.94, 1.07]$ } \\
Asia * Platform Tenure 
    & \shortstack{0.97 \\ $[0.91, 1.03]$ } 
    & \shortstack{1.00 \\ $[0.93, 1.07]$ } 
    & \shortstack{1.12 \\ $[0.99, 1.27]$ } 
    & \shortstack{0.95 \\ $[0.88, 1.03]$ } \\
\end{tabularx}

\label{tab:reg-behaviors}
\end{table*}

\clearpage

\end{document}